\documentclass[aps,twocolumn,superscriptaddress,nobibnotes]{revtex4-1}

\usepackage{graphicx}
\usepackage{amsmath}

\begin{document}


\title{Mitigation of multipacting in 113~MHz superconducting RF photo-injector}


\author{I.~Petrushina}
\email{ipetrushina@bnl.gov}
\affiliation{Department of Physics and Astronomy, Stony Brook University, Stony Brook, NY 11794, USA}
\affiliation{Collider-Accelerator Department, Brookhaven National Laboratory, Upton, NY 11973, USA}
\author{V.N.~Litvinenko}
\affiliation{Department of Physics and Astronomy, Stony Brook University, Stony Brook, NY 11794, USA}
\affiliation{Collider-Accelerator Department, Brookhaven National Laboratory, Upton, NY 11973, USA}
\author{I.~Pinayev} 
\author{K.~Smith}
\author{G.~Narayan}
\author{F.~Severino}
\affiliation{Collider-Accelerator Department, Brookhaven National Laboratory, Upton, NY 11973, USA}

\date{\today}

\begin{abstract}
Superconducting RF (SRF) photo-injectors are one of the most promising devices for generating continuous wave (CW) electron beams with record high brightness. Ultra-high vacuum of SRF guns provides for long lifetime of the high quantum efficiency (QE) photocathodes, while SRF technology provides for high accelerating gradients exceeding 10~MV/m. It is especially true for low frequency SRF guns where electrons are generated at photocathodes at the crest of accelerating voltage. Two main physics challenges of SRF guns are their compatibility with high QE photocathodes and multipacting. The first is related to a possibility of deposition of photocathode materials (such as Cs) on the walls of the SRF cavity, which can result in increased dark current via reduction of the bulk Nb work function and in enhancing of a secondary electron emission yield (SEY). SEY plays critical role in multipacting (e.g. an exponential growth of the multipactor discharge), which could both spoil the gun vacuum and speed up the deposition of the cathode material on the walls of the SRF cavity. In short, the multipactor behavior in superconducting accelerating units must be well understood for successful operation of an SRF photo-injector.
In this paper we present our studies of 1.2~MV 113~MHz quarter-wave SRF photo-injector serving as a source of electron beam for the Coherent electron Cooling experiment (CeC) at BNL. During three years of operating our SRF gun we encountered a number of multipacting zones. We also observed that presence of $\textrm{CsK}_{2}\textrm{Sb}$ photocathode in the gun could create additional multipacting barriers. We had conducted a comprehensive numerical and experimental study of the multipactor discharge in our SRF gun. We had developed a process of crossing the multipacting barriers from zero to the operational voltage without affecting the lifetime of our photocathode and enhancing the strength of multipacting barriers. We found a good agreement between the results of simulations and our experimental data.
\end{abstract}


\maketitle

\section{Introduction}
Superconducting radio-frequency (SRF) electron guns are frequently considered to be the favorite pathway for generating the high-quality, high-current beams needed for the future high-power energy-recovery linacs. SRF guns can find unique scientific and industrial applications, such as driving high-power X-ray and extreme ultraviolet (EUV) CW free electron lasers (FELs) \cite{1,2,3,4,5,6,7,8,9,10}, intense $\gamma$-ray sources \cite{11,12,13,14}, coolers for hadron beams \cite{15,16,17,18}, and electron-hadron colliders \cite{19,20,21}. The quality of the generated electron beam---both its intensity and brightness---is extremely important for many of these applications. 

SRF technology is well suited for generating CW electron beams in high accelerating gradient environments. This is especially true for low-frequency, 100~MHz scale, quarter-wave SRF guns, where, in contrast with high frequency CW SRF guns \cite{22,23}, the electrons are generated at the photocathode near the crest of the accelerating voltage. To a degree, such SRF guns are similar to DC guns but offer both high accelerating gradient and higher beam energy at the gun exit. 

Still, there are a number of challenges associated with developing and operating SRF photo-electron guns. One such challenge is the most common problem of any RF cavity:  multipacting (MP)---a resonant process in which an electron avalanche builds up within a small region of a cavity surface. The MP electron absorbs RF power and deposits it as heat in the cavity walls. Hence, MP could either prevent the cavity from reaching the designed accelerating voltage, or deposit excessive heat into the wall of the cavity and, in the case of an SRF cavity, cause it to quench. 

In order for multipactor discharge to develop, several resonant conditions have to be satisfied. Multipacting usually starts with a few (or even a single) primary electron(s) being present inside the cavity. Such electrons can appear when a cosmic X-ray strikes a molecule of residual gas or the cavity wall, or as a result of dark current/field emission. These electrons, if in the accelerating phase of the RF fields, gain energy and hit a cavity wall. Depending on the energy of the primary electrons and the material of the cavity walls (surface), more than one secondary electron can be emitted. If these secondary electrons are released in the accelerating phase of the electric field, and the local geometry of the cavity is such that the particles can reach the surface again and return in the same phase, the process repeats. This periodic enhancement of the electron population will lead to the formation of an electron avalanche, e.g. the multipacting. The number of secondary electrons being released is called Secondary Electron Yield (SEY) and it strongly depends on the material of the wall's surface, the energy of the primary electron $E_{i}$ and the impact angle of the electrons. Primary electrons with a low initial energy of about a few eV generate secondary electrons near the surface of the wall (e.g. within a small depth) with high probability for them to escape. In contrast, electrons with higher energy penetrate deep into the bulk of the wall and the probability of the secondary electron emission is greatly reduced. This explains the fact that multipacting is usually localized in areas with low electric fields and frequently occurs at low levels of accelerating gradient. The SEY is material-dependent and it is strongly affected by the quality of the surface and the treatment of the cavity. Since all interactions occur at small depth, the SEY is very sensitive to any adsorbents in the cavity walls. Specifically, we believe that any deposition of materials from high QE photo-cathodes, especially Cs, would increase the SEY when compared with bulk Nb of SRF cavity walls or Cu structures supporting the cathode insertions.

Being fairly complicated phenomena, the multipactor discharge must be well understood in order to achieve desirable performance of any accelerating RF device, and especially SRF photo-injectors. Otherwise, MP barriers can prevent an SRF gun from reaching the operating voltage by absorbing all available RF power and capturing the gun at one of the strong MP levels or/and generating significant outgassing from the cavity walls and poisoning the photocathode. We noticed experimentally that when the gun is captured at a MP level for an extended period of time, e.g. few minutes, the MP barriers are becoming stronger. In other words, it makes MP zones impassable with the full coupling and RF power. The only remedy we found was to let the gun rest from 30 minutes to 4-8 hours. We attribute this behavior to a possible increase of SEY caused by prolonged multipacting. We frequently observed such behavior of our SRF gun during the first year of its operation. Better understanding of the multipacting and processes in the gun affecting SEY allowed us to develop a process of crossing the MP barrier without adverse effects on the SRF gun performance.

In this paper we present a complete study of multipacting phenomena in our SRF photo-injector built for the CeC experiment at BNL \cite{15,24,25,26,27,28}. We start with the first observations of multipacting during the earlier commissioning stages, where it was a serious limitation to the system performance. We continue with describing the series of numerical simulations investigating areas of the cavity affected by the MP discharge and understanding of the system behavior. Finally, we present our latest commissioning results, where we were able to successfully overcome the multipactor discharge problem and achieve a stable operational regime of our SRF gun.

\section{SRF QW Photoelectron Gun}

The superconducting 113~MHz photo-injector based on Quarter Wave Resonator (QWR) was designed to serve as a source of electron beam for the Coherent electron Cooling \cite{15} Proof-of-Principal (CeC PoP) experiment \cite{25}. The goal of the experiment is to demonstrate efficient cooling of a hadron bunch circulating in Relativistic Heavy Ion Collider (RHIC) using this novel technique. The CeC system is currently undergoing commissioning at Brookhaven National Laboratory \cite{pinayev2016commissioning}.

\begin{figure}[h!]
 	\includegraphics[width=1\linewidth]{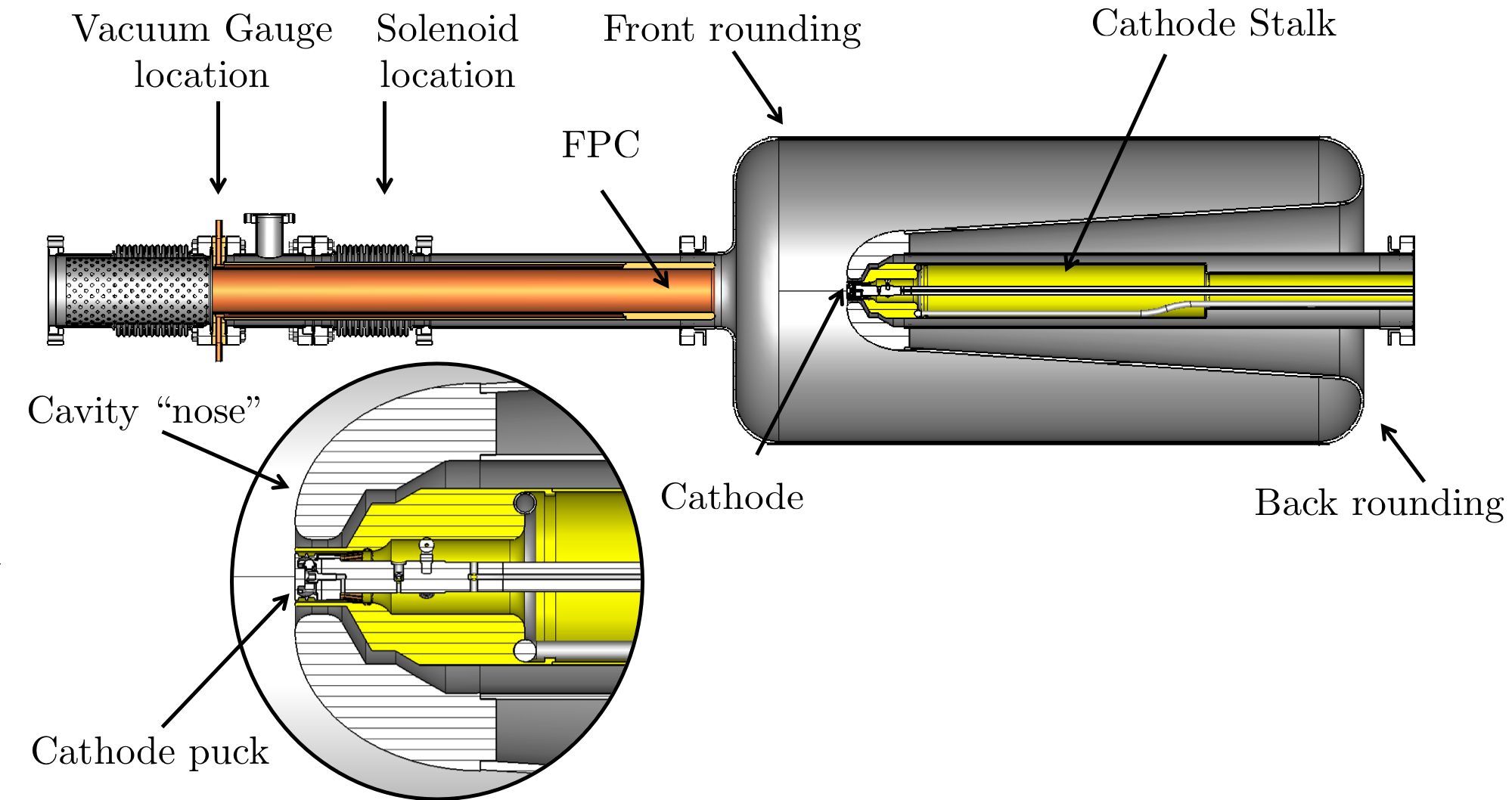}
 	\caption{Simplified geometry of the SRF Gun.\label{fig:Gun}}
 \end{figure}
 
The gun was built by Niowave Inc. in 2014 for the CeC experiment as a modification of their QWR accelerating cavity. The cathode stalk was built at Stony Brook University as a part of the photocathode development R\&D. A simplified model of the gun is shown in Fig.~\ref{fig:Gun}. The gun operates at 1.05~MV of accelerating voltage and generates electron bunches with a charge up to 4~nC and repetition rate of 78~kHz. The main RF parameters of the gun are shown in Table~\ref{tab:Gun}. The fundamental power coupler (FPC) serves two functions: it couples power into the SRF cavity of the gun from a 4~kW RF transmitter and provides for a fine tuning of the gun's resonant frequency. The $\textrm{CsK}_{2}\textrm{Sb}$ photo-cathode deposited on a molybdenum puck operates at room temperature, while the gun cavity operates at liquid helium temperature of 4~K. The cathodes are inserted into the gun using an ultra-high vacuum (UHV) transport system into a hollow stainless steel cathode stalk, which is coated with 100~micron thick layers of copper and gold, to reduce heat emission into the 4~K system and is kept at room temperature by a circulating water. The stalk also serves as a half-wave RF choke with a pick-up antenna located outside of the gun cryostat. 

\begin{table}[h!]
	\caption{\label{tab:Gun}RF parameters of the gun.}
	\begin{ruledtabular}
		\begin{tabular}{lc}
			Parameter & Value\\ \hline
			Frequency, MHz & 113\\
			Quality Factor w/o cathode & $3.5\times10^9$\\
			$R/Q$, $\Omega$ & 126\\
			Geometry Factor, $\Omega$ & 38.2\\
			Operating temperature, K & 4\\
			Accelerating voltage, MV & 1.05-1.2\\
		\end{tabular}
	\end{ruledtabular}
\end{table}
 
The CeC SRF accelerator, including the SRF gun, uses liquid helium from RHIC refrigerator, which operates only when RHIC is running. Hence, the SRF gun must be operated only during the RHIC runs (typically from February through June), and hence the operation periods for our gun are measured in RHIC runs. The CeC gun generated its first beam in spring of 2015 \cite{29} but was fully commissioned and characterized during the RHIC runs in 2016 and 2017. The gun demonstrated accelerating voltage up to 1.6~MV in a pulse mode and 1.25~MV in CW mode, but operating above 1.15~MV usually results in excessive X-ray radiation and increases dark current. Hence it is regularly operated at accelerating voltage of $1.1\pm 0.05$ MV.

During the gun commissioning in 2016 we encountered a number of multipacting barriers in the SRF gun \cite{xin2015experimental,28}. The multipacting could be distinguished by an observation of the significant vacuum activity, inability of the gun to reach required voltage and/or by a noise in the low-level RF (LLRF) loops. We observed that the magnetic field of the gun solenoid led to substantial vacuum activity in the FPC (especially for a field of about 400~Gs). This solenoid is the first focusing element of the lattice, and it generates a magnetic field at the bellow section of the FPC. The complicated geometry of the bellows is known for creating resonant conditions for the stable multipacting trajectories. This multipacting was happening at very low RF voltages (few kVs) and we found a simple solution for this problem by turning the gun solenoid on after reaching the gun's operational voltage.

The most stubborn multipacting barrier was found at about 40~kV of accelerating voltage. It is our understanding that this is a strong first-order multipacting occurring in the front rounding of the cavity. In order to overcome this barrier we had to insert the FPC into the position with maximum possible coupling and use the maximum power of our transmitter, which was limited to 2~kW during the run 2016. As soon as the gun passed above 100~kV voltage, there were no problems with reaching the operating voltage by adjusting coupling via FPC position in a phase-lock loop mode.

\begin{figure}[h!]
\includegraphics[width=1\linewidth]{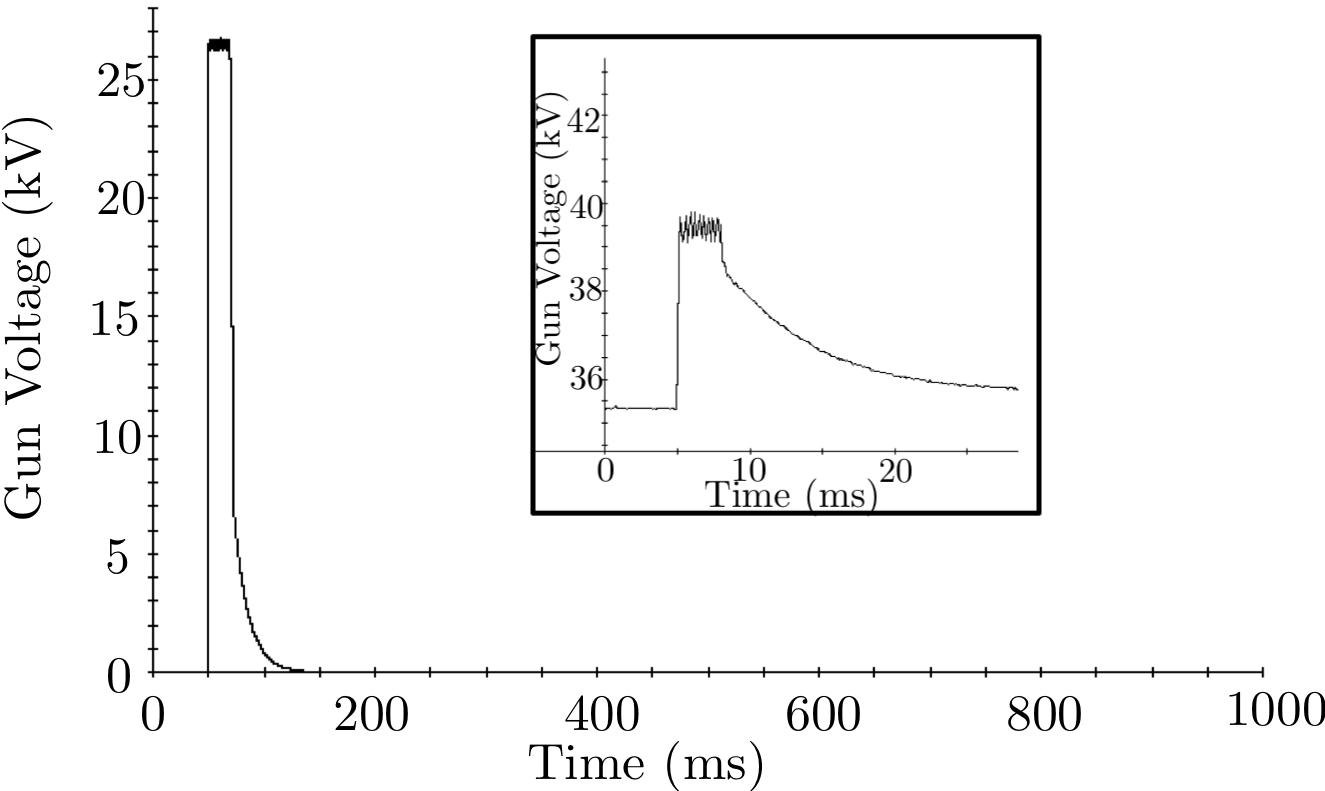}
\caption{Typical multipacting events at 26~kV and 40~kV of accelerating voltage during RHIC run 2016. RF pulse with 2~kW peak power was caught by multipacting barrier with typical zigzag (noisy) pattern in the read-back voltage, followed by an exponential decay after  the RF pulse went off. Sampling rate of the data logging is 1~kHz, which could be insufficient to estimate the period of MP oscillations.}
\label{fig:Experiment}
\end{figure}

We also had found that inserting the $\textrm{CsK}_{2}\textrm{Sb}$ photocathode as well as exposing the gun to 40~kV multipacting would aggravate the strength of the multipacting barrier. A plausible explanation of this phenomenon is that multipacting initiates deposition of a high SEY material from the photocathode surface to the wall of the gun cavity and enhances the process of multipacting. By the end of run 2016 it became challenging, if not impossible, to overcome this level. After extracting the cathode from the cavity, the gun reached the operational voltage without any problems, which confirmed our suspicion, that  a high SEY material of the photocathode is responsible for the enhanced multipacting strength. 
 
In addition to the challenge of reaching the operating voltage, the multipacting was generating serious vacuum excursions, which were very damaging to the photocathode quantum efficiency (QE). Our gun is equipped with a cathode manipulation and storage system (simply called the ``garage"), which can house three cathode pucks. The $\textrm{CsK}_{2}\textrm{Sb}$ photocathodes were produced with a rather high QE of 2-8\% and a number of them maintained high QE when they were inserted into the gun. However, exposure to the sustained multipacting at the 40~kV barrier for a few minutes would deteriorate the QE to the level of 0.01\%-0.1\%. Since the 4~K walls of the gun cavity work as a cryogenic vacuum pump, it is conceivable that the QE drop during the multipacting event was also associated with a bombardment by stray electrons originating from the multipacting zone and reaching the cathode. It also could explain the removal of Cs from the cathode surface and its consequent deposition onto the cavity walls. 

In addition to the 40~kV multipacting zone, we observed similar zones at about 26-27~kV, 19~kV and sometimes as low as 2~kV (see Fig.~\ref{fig:Experiment}). We undertook dedicated efforts to better understand multipacting in our gun and to develop necessary methods and codes leading to successful gun operation during the run 2017 with high QE photocathodes operating for months without degradation.

\section{Simulations}

As mentioned above, there are several resonant conditions that must be satisfied simultaneously for multipacting to occur. This makes multipacting analysis and simulations a fairly complicated problem. One has to take into account the detailed geometry of the gun and SEY coefficients (with energy dependencies) for all its materials. We have performed a series of multipacting simulations using the sophisticated 3D codes---CST Studio and ACE3P. This comprehensive analysis of the EM fields and multipacting in the gun provided us with the necessary information about the multipacting zones and their relative strengths. 

\subsection{CST STUDIO}

CTS Studio is one of the most commonly used tools for the 3D design of accelerating structures, which provides solvers for various aspects of RF simulations with a user-friendly interface \cite{cst}. For multipacting simulations CST utilizes a hexahedral mesh with a perfect boundary approximation to avoid the stair-step mesh. Such meshing, when compared to a tetrahedral one, has a disadvantage with obtaining an accurate representation of small features in sophisticated RF systems (such as SRF gun) with modest computational resources.

 \begin{figure}[h!]
	\includegraphics[width=1\linewidth]{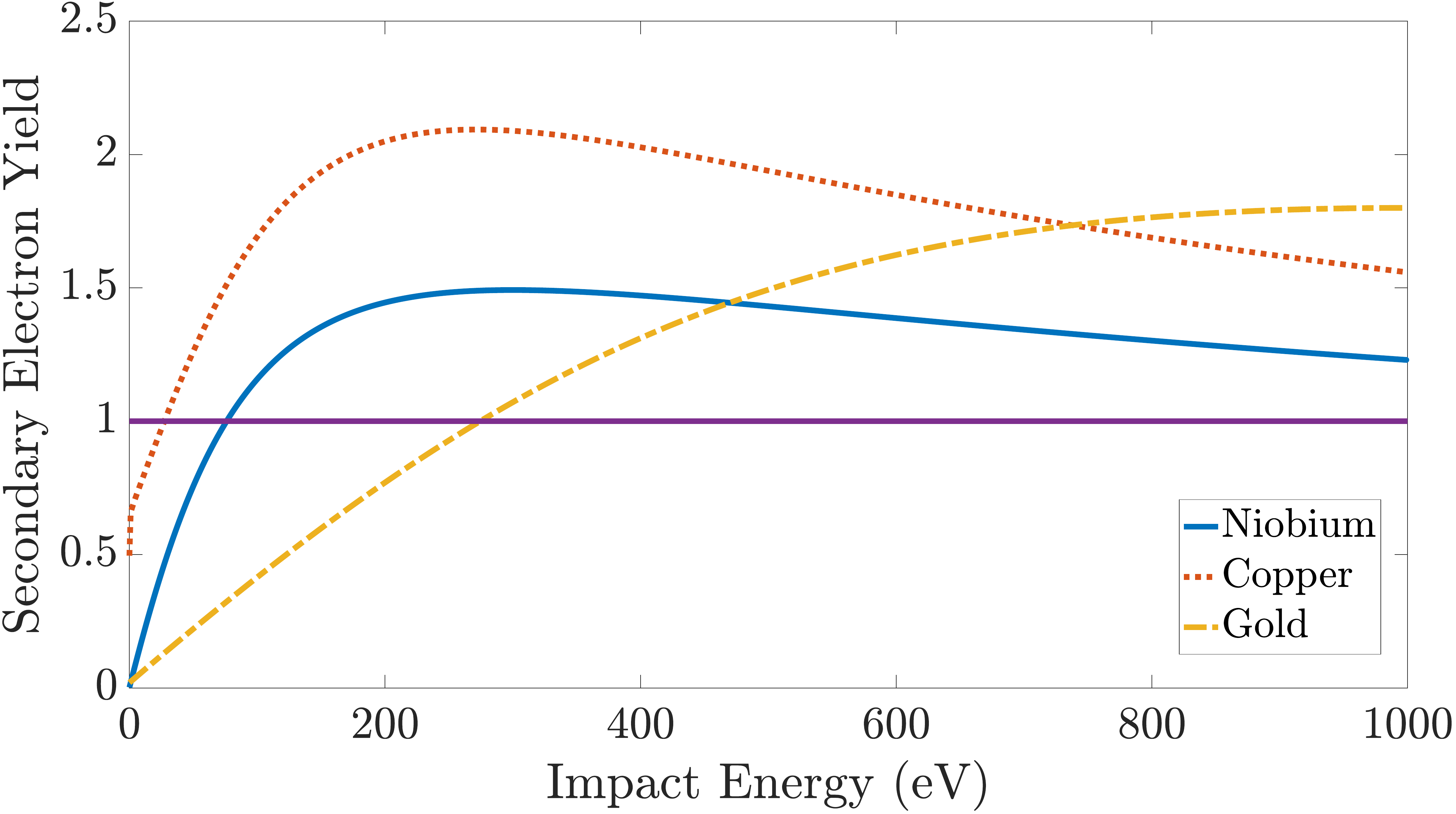}
	\caption{Secondary Emission Yield as a function of primary energy of electrons for the materials used in the SRF gun---gold, niobium and copper.\label{fig_mat}}
 \end{figure}

The electromagnetic field distribution in the gun was first calculated using CST Microwave Studio's Eigenmode Solver. A hexahedral mesh of 7.1~million meshcells per quarter of the geometry (280~meshcells per wavelength) was generated in order to compute an accurate distribution of electromagnetic fields within the cavity volume. Small geometrical features could create a resonant condition for the multipacting, hence they had to be considered with a higher precision. Therefore, the mesh was enhanced in the cathode area and in the area of the two bellows. The resulting fields were imported to CST Particle Studio which is able to perform multipacting simulations using the Particle In Cell (PIC) solver. 
 
Primary electrons were emitted from the whole surface of the FPC, stalk and cathode, and also from a thin $15^{\circ}$ slice of the cavity surface. The latter was allowed by the axial symmetry of the gun and significantly reduced required computational resources. Initial electrons were emitted with random energies from 1 to 5~eV  and random phases during the first RF cycle. Then the simulations were performed for a total of 40~RF periods.

The SEY properties used for the simulation were uploaded from the CST Material Library and are shown in Fig.~\ref{fig_mat}. It is important to mention that CST uses the Furman-Pivi model of the secondary electron emission \cite{Furman_Pivi} and takes into account the dependence of SEY on the incident angle of electrons.

\begin{figure}[h!]
	\includegraphics[width=1\linewidth]{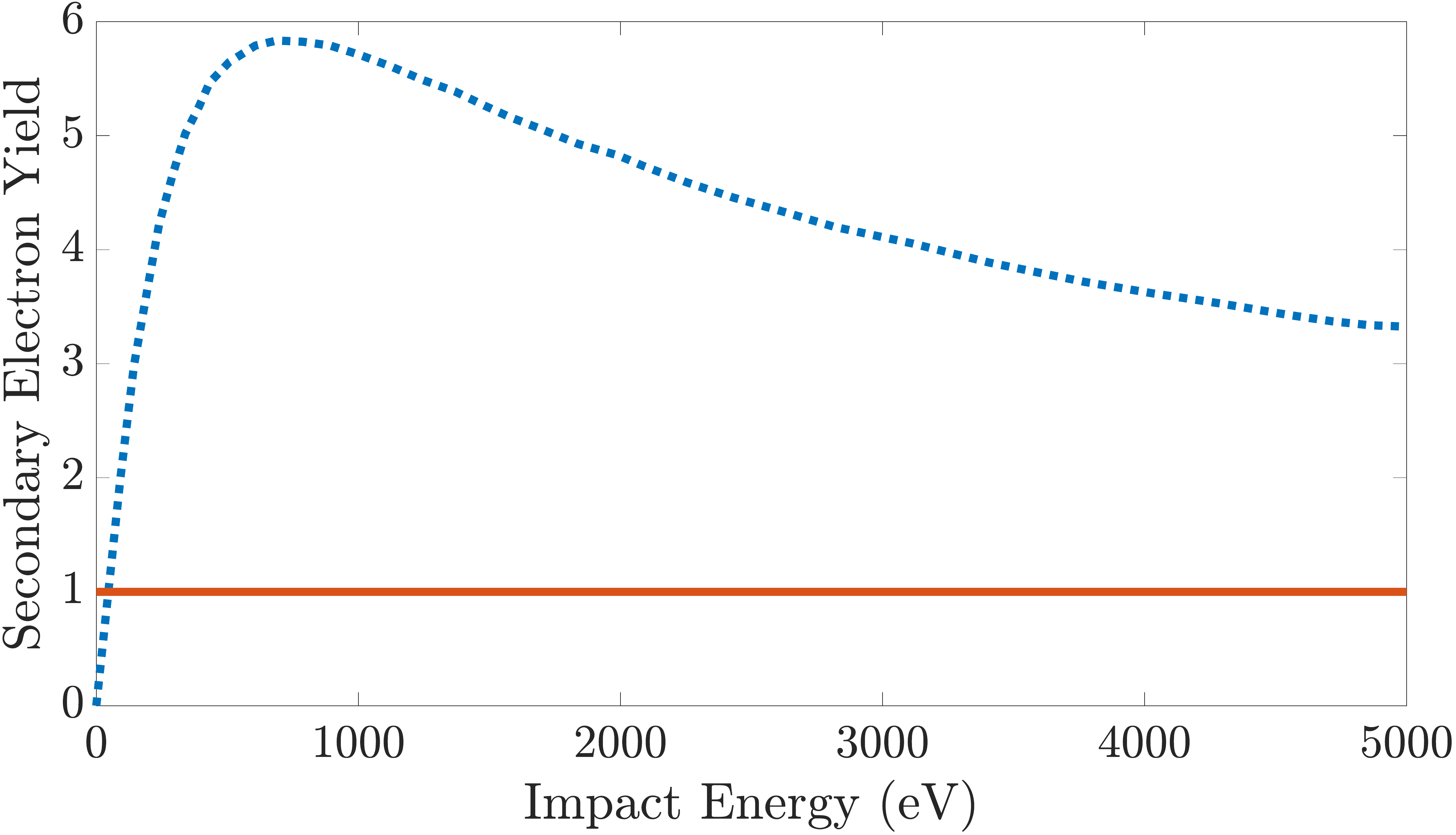}
	\caption{Secondary Emission Yield of $\textrm{Cs}_{3}\textrm{Sb}$.\label{fig_cath}}
\end{figure}

Since the actual SEY of the exact alkali antimonide compound of our photo-cathode is unknown, it was decided to utilize properties of $\textrm{Cs}_{3}\textrm{Sb}$ for the following simulations (see Fig.~\ref{fig_cath}) \cite{michalke1993photocathodes}.

Exponential growth of the secondary electrons within the cavity body is one of the signs of multipactor discharge. During the simulation we monitored the number of electrons in the cavity as a function of time and also observed locations of stable multipacting trajectories. The simulations were first performed without external solenoidal field. As we mentioned above, calculations for such a complicated design require a high density mesh. This leads to a very time consuming simulation, especially if we are analyzing an electron avalanche formation for a variety of accelerating field levels in the gun.  Hence, first we focused our simulations on studying of the field levels at which multipacting behavior was observed experimentally. 

\begin{figure}[h!]
	\includegraphics[width=1\linewidth]{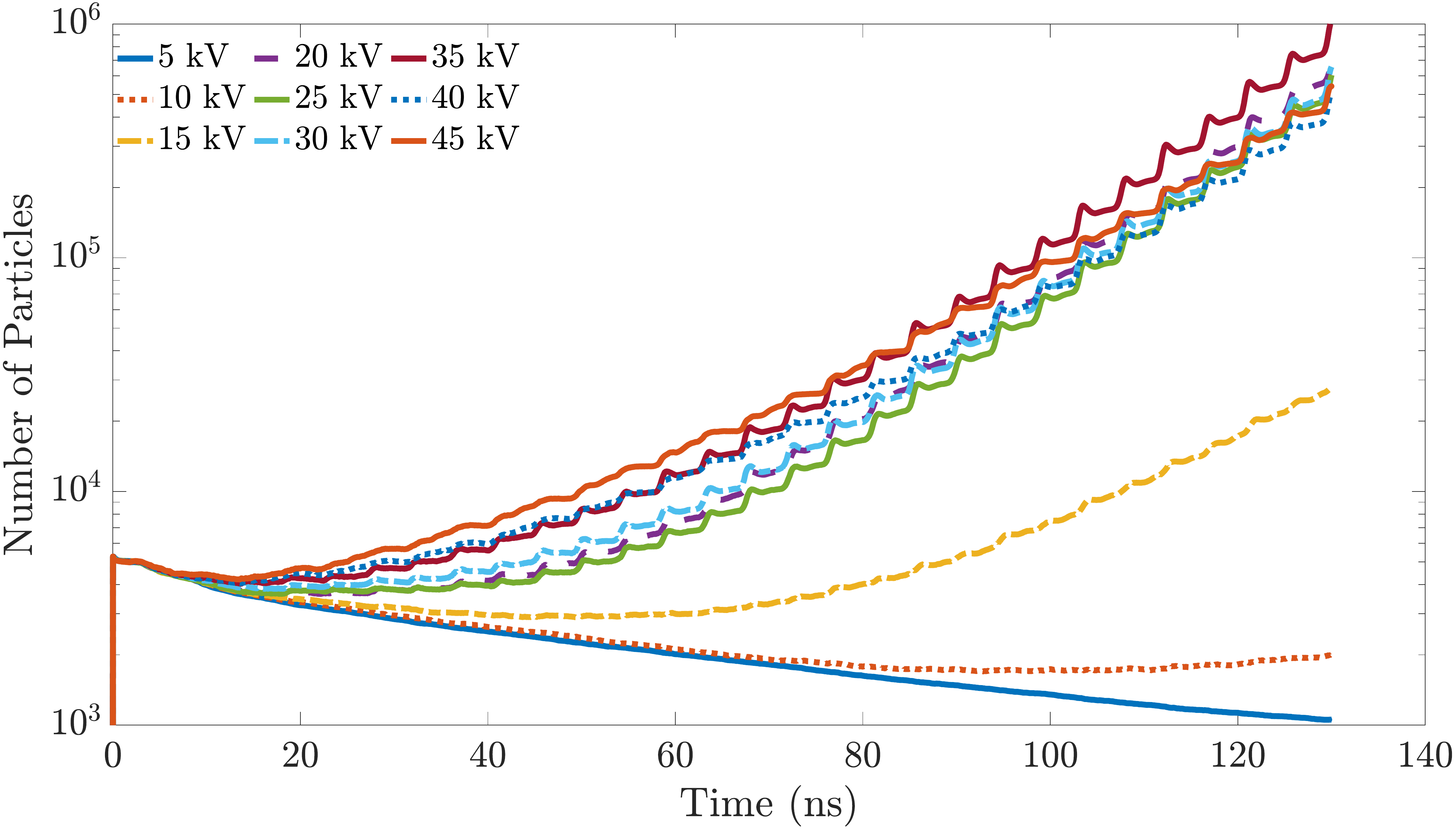}
	\caption{Number of particles within the cavity as a function of time.\label{fig_cst}}
\end{figure}

\begin{figure*}[!hbt]
	\centering
	\includegraphics*[width=\textwidth]{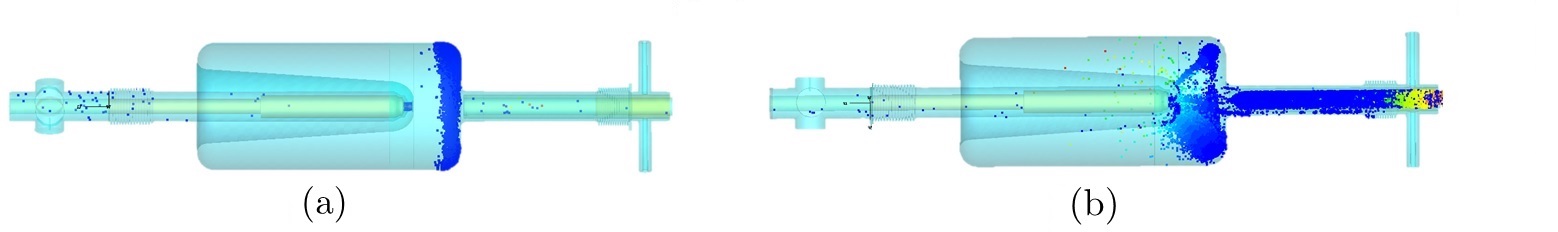}
	\caption{Stable trajectories in the gun based on the CST simulation results: (a) trajectories at the front rounding of the cavity at 28~kV of accelerating voltage, (b) stable trajectories between the nose and the front rounding of the cavity at 40~kV of accelerating voltage.}
	\label{CST_trajectories}
\end{figure*} 

Our simulations indicated a strong exponential growth of secondary electrons at voltages above 20~kV up to 45~kV (see Fig.~\ref{fig_cst}), which corresponds to the MP barriers most frequently encountered during the commissioning. Stable trajectories were concentrated in the front rounding of the cavity, as it was predicted earlier (see Fig.~\ref{CST_trajectories}a). We also observed several trajectories between the nose and the front rounding of the cavity (Fig.~\ref{CST_trajectories}b), which could be harmful for the surface of the photo-cathode, but such trajectories did not survive more than a few RF cycles, and were not considered to be dangerous. Stable multipacting trajectories were observed in the gap of the FPC and the bellow as well.

\begin{figure}[h!]
	\includegraphics[width=1\linewidth]{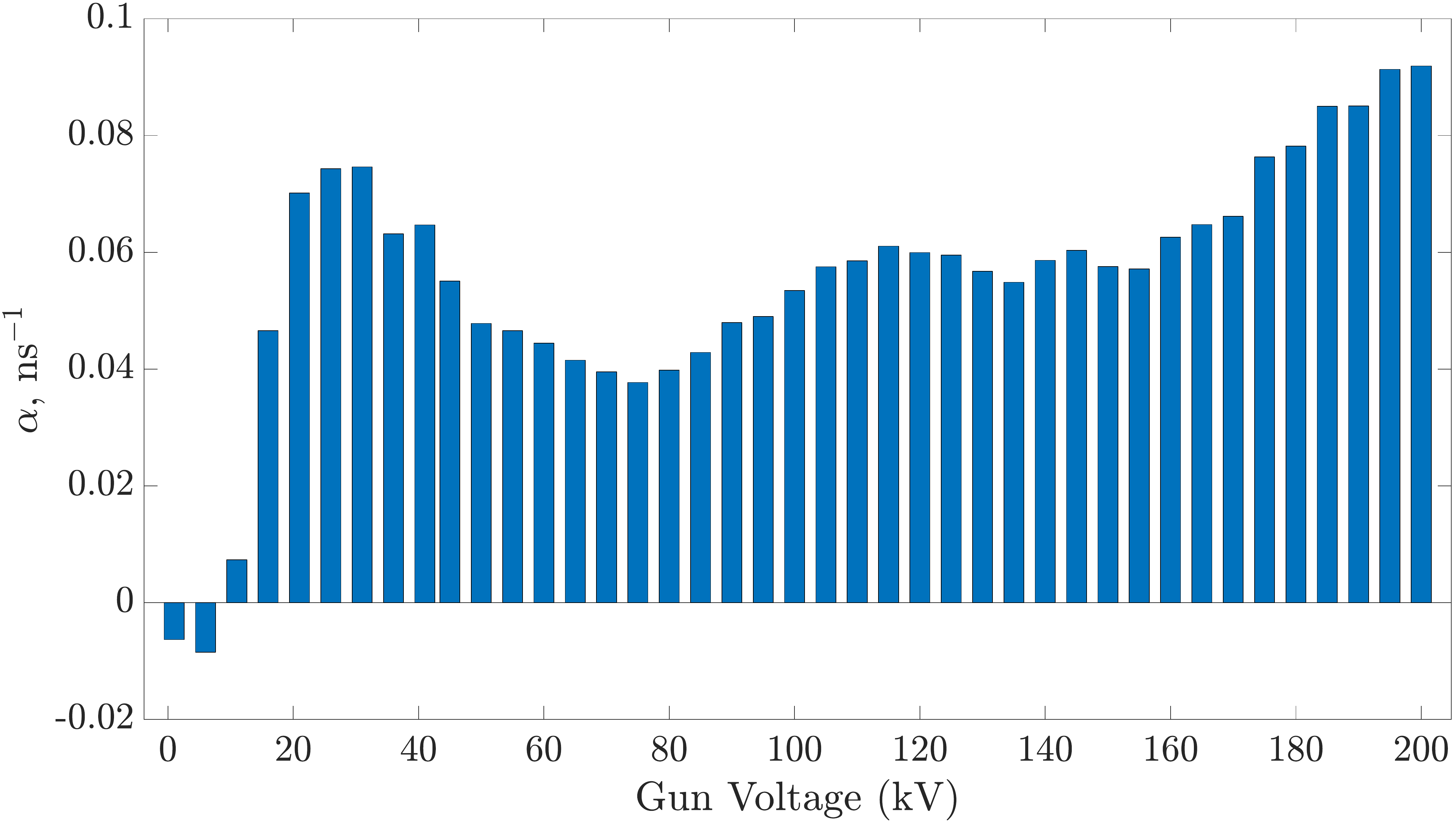}
	\caption{Exponential growth rate of electron avalanche $\alpha$ as a function of accelerating voltage in the SRF gun simulated by CST PS.\label{fig_alpha}}
\end{figure}

An established electron avalanche can be described by an exponential growth:

\begin{equation}
N_{e}(t)=N_{0}e^{\alpha t},
\label{eq:Nexp}
\end{equation}

where $N_{0}$---number of primary electrons, and $\alpha$ is the exponential growth rate representing the strength of the multipacting barrier. Negative values of $\alpha$ would indicate the decaying number of electrons and absence of a multipacting barrier. According to the simulation results using CST PS in a wide range of accelerating voltage (see Fig.~\ref{fig_alpha}), positive $\alpha$ was observed at the gun voltages above 10~kV with prominent peaks around 30~kV, 110~kV and 200~kV, which is in a good agreement with the experimental observations. Stable multipacting trajectories were observed at the front rounding of the cavity, FPC gap and the back rounding of the gun correspondingly.

\begin{figure}[h!]
	\includegraphics[width=1\linewidth]{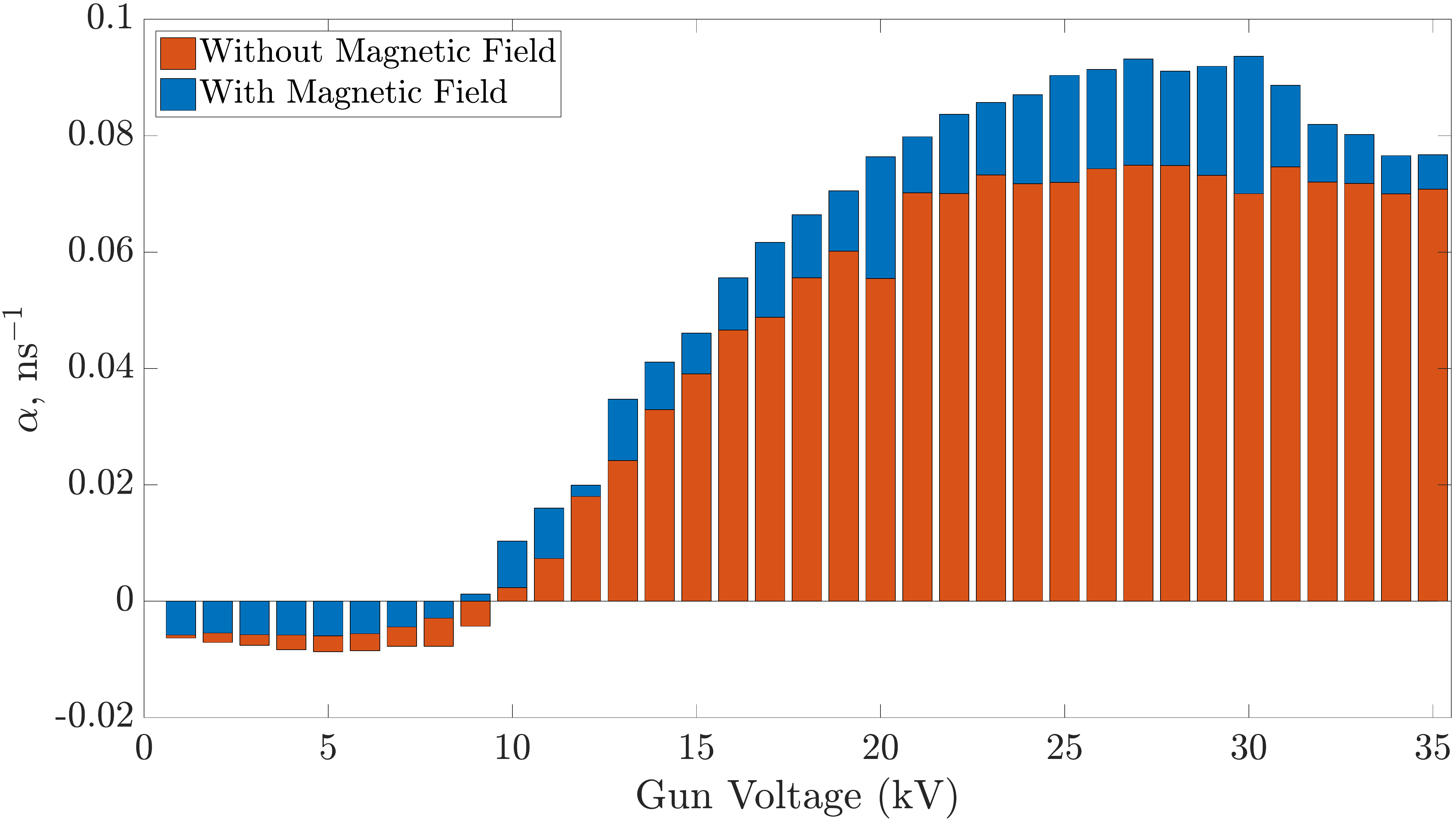}
	\caption{Exponential growth rate of electron avalanche $\alpha$ as a function of accelerating voltage in the SRF gun simulated by CST PS with (blue) and without (orange) external solenoidal field.\label{fig_alphaB}}
\end{figure}

We performed similar simulations taking into account the magnetic field of the solenoid with a peak field of 400~Gs (with the field map simulated by CST Electromagnetic Studio) at low levels of accelerating voltages where we experimentally observed multipacting. Figure~\ref{fig_alphaB} demonstrates that according to the simulation, external magnetic field increases the strength of multipacting, and leads to a shift of MP barriers to lower levels of accelerating voltage which agrees with experimental results.

In order to extend our simulation to a full range of operational voltage of our gun and study stable MP trajectories in more details, we needed to find a code allowing the use of a very refined mesh without requiring outrageous computer resources.

\subsection{ACE3P}

Advanced Computational Electromagnetics 3D Parallel suite (ACE3P) is a 3D parallel finite element code for design and development of accelerating units which includes packages for various aspects of RF related problems such as electromagnetic (Omega3P), thermal and mechanical effects (TEM3P) \cite{ACE3P}. Track3P is designated for multipacting and dark current simulations \cite{Track3P,30}. All of the calculations performed with ACE3P utilize massively parallel computers which allows for increased memory (problem size) and speed of the simulation \cite{ge2015advances}.

The geometry of the gun was split into several volumes for separate meshing of each part, which made it possible to apply a fine geometry-adaptive mesh with a better mesh resolution in the areas of the bellows and the cathode. The total number of meshcells in the cavity used for simulations was about 1.1~million tetrahedrons.

The electromagnetic field distributions were computed using Omega3P and then used in Track3p for multipacting simulations.  

Since the areas of the cavity potentially affected by the multipacting discharge were already known, it was easier to perform simulations for each area separately. First, primary electrons were set to be emitted from a thin slice of the cavity wall, similarly to what was done before using CST PS.  Initial electrons escaped the surface of the cavity with energy of 2~eV during the first RF cycle, while the simulation was performed for total of 50~RF periods. Initial velocities of all primary particles were perpendicular to the cavity surface, and there was no spread in emission energy. Additionally, it is important to mention that Track3P does not include space charge effects in multipacting simulations, and, unlike CST, does not take into account dependence of the SEY on the incident angle of electrons.

Since the software can utilize only two materials with different SEY in the simulation at once, it was convenient to use SEY of copper for both the FPC and the cavity walls, which would take into account that the actual quality of the cavity surface is supposedly worse than that of ideal niobium (see Fig.~\ref{fig_mat}). The SEY of the cathode used in the simulation is shown in Fig.~\ref{fig_cath}. Since the main area of concern in this calculation was the cavity itself, the magnetic field of the solenoid was not included in this setup.

As a result of simulation, Track3P generates a large amount of particles (the code increases the number of secondary electrons after every impact according to the corresponding SEY curve) and fields, provides information about parameters of stable multipacting trajectories (location, energy of impact, etc.), determines areas of the geometry affected by multipactor discharge, and calculates the Enhancement Counter (EC). EC is usually defined as follows:

\begin{equation}
EC=\delta_{1}\times\delta_{2}\times...\times\delta_{n},
\end{equation}

where $\delta_{1}, \delta_{2},...,\delta_{n}$---number of secondary electrons emitted after the $1^{st},2^{nd},...,n^{th}$ impact respectively, and is directly determined by the SEY of the surface material. Usually, the sign of possible multipacting is when $EC$ exceeds unity, and the higher the $EC$ value, the more complicated it can be to process multipacting during conditioning. By evaluating $EC$ in the whole range of operational voltage of an RF device, one can determine multipacting zones, and study stable trajectories at these levels in more detail.

It is important to mention that during the run 2016, the cathode puck was aligned with the cavity nose, but in order to provide additional primary focusing to the beam, it was decided to start operating the gun with a slight, about 6 mm, recess of the cathode surface relative to the cavity nose. Such a change of the geometry causes a slightly different electromagnetic field distribution, which can lead to the appearance of new resonant conditions for multipacting. However, simulations showed that the recess of the cathode doesn't change multipacting behavior in the gun dramatically, and shifts multipacting barriers by approximately 1~kV towards lower accelerating voltages. The following results were performed for the geometry with the cathode puck being recessed by 6 mm, since it would be more applicable for the future gun operation.

\begin{figure}[h!]
	\includegraphics[width=1.1\linewidth]{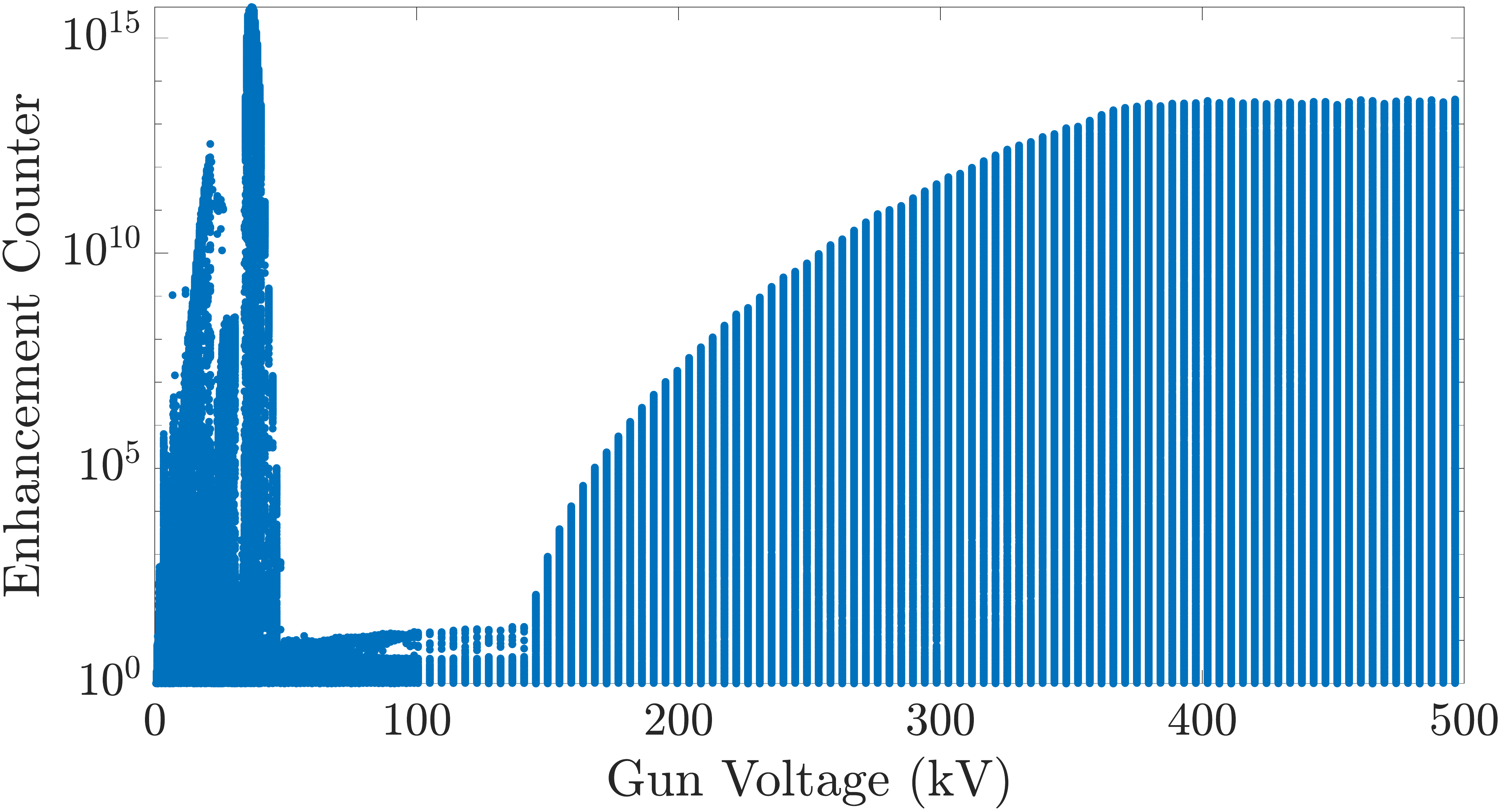}
	\caption{Enhancement Counter as a function of accelerating voltage in the SRF gun simulated by Track3P.\label{fig:EC_Track3P}}
\end{figure}

The resulting $EC$ for the cavity (see Fig.~\ref{fig:EC_Track3P}) showed substantial increase in a number of particles at low voltages---the first peaks on the graph correspond to 30~kV and 40~kV, which agrees with multipacting bands observed during the commissioning and predicted by CST PS.

After the possible multipacting bands in the cavity were determined from the $EC$ function, it was necessary to perform a detailed study of resonant trajectories. It is important to determine the location of stable trajectories, their impact energies, order, and for how many RF cycles the trajectory can survive. Order of a multipactor is the number of RF cycles it takes for an electron to travel between two subsequent impacts. 

\begin{figure}[h!]
	\includegraphics[width=1\linewidth]{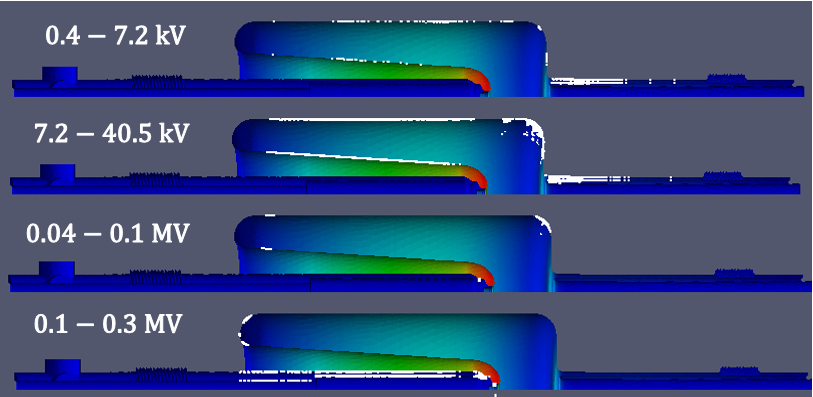}
	\caption{Areas of the gun affected by multipacting (shown in white). Primary electrons are emitted from the surface of the cavity.\label{traj1}}
\end{figure}

Figure~\ref{traj1} shows the areas of the gun surface affected by MP at different levels of accelerating voltage. It was observed that even though the primary particles were emitted from the surface of the cavity itself, the stable trajectories at low levels of the voltage (0.4-40.5~kV) were found to be in the FPC gap. Also, at low voltages, stable trajectories were present within the cavity body between the inner and outer conductors of the quarter-wave resonator. Those trajectories moved from the nose of the cavity toward the back rounding of it, when the voltage was increasing, and survived less than 20~RF cycles.

The trajectories corresponding to the peaks of the EC function at 30~kV and 40~kV are located in the front rounding of the cavity, as was already determined. These trajectories are mainly $1^{st}$ and $2^{nd}$ order MP trajectories which survive more than 50~RF cycles, and can be more challenging to supress than higher order multipacting. Higher order MP trajectories tend to be less stable and more susceptable to the space charge, which makes them easier to process if allowed by geometrical conditions. For voltages higher than 100~kV, the trajectories move toward the back rounding of the cavity and become stable $1^{st}$ order MP trajectories at about 200~kV.

\begin{figure*}[!hbt]
	\centering
	\includegraphics*[width=\textwidth]{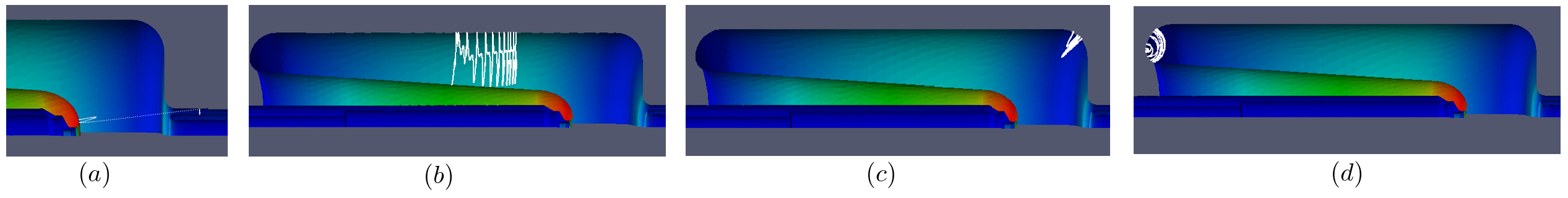}
	\caption{Stable MP trajectories in the gun (shown in white). (a) -- $1^{st}$ order MP in the FPC at 7~kV, impact energy $E_{i}=28$ eV; (b) -- $8^{th}$ order MP in the cavity at 7~kV, $E_{i}=900$ eV; (c) -- $1^{st}$ order MP in the front rounding of the cavity at 40~kV ($E_{i}=500$ eV); (d) -- $1^{st}$ order MP in the back rounding of the cavity at 200~kV, $E_{i}=20$ eV.}
	\label{trajectories}
\end{figure*}

One can see from Fig.~\ref{trajectories}a, that simulation showed certain trajectories in the gap between the nose of the cavity and its front plate, but after close consideration it was found that after 1 RF cycle, secondary electrons get accelerated by the electric field, move towards the FPC, and become $1^{st}$ order multipacting trajectories in the gap between the FPC and outer conductor.

At accelerating voltages above 150~kV, multipacting trajectories were found to be in the gap between the stalk and the cavity wall. These are stable $1^{st}$ and $2^{nd}$ order trajectories, which first appear in the smaller gap, and move towards the larger gap of the RF choke at higher voltages.

Since the gold plated stalk is located inside the niobium tube, which creates geometrical conditions for multipacting, threshold levels of voltage within this gap can be estimated analytically using a two plane approach. Considering the fields within the gap uniformly distributed, it can be treated as a two-point multipacting between two plates. According to \cite{wangler2008rf}, conditions for two-point multipacting between two plane surfaces can occur at the range of voltages between $V_{min}$ and $V_{max}$, which can be estimated as follows:

\begin{equation}
V_{max}=\frac{m\omega^2 x^2}{2e},
\end{equation}

where $x$---distance between the surfaces, $\omega$---resonant frequency, $m, e$---mass and charge of an electron.

\begin{equation}
V_{min}=\frac{m\omega^2 x^2}{e}\frac{1}{\sqrt{4+(2n+1)^2\pi^2}},
\end{equation}

where $n$ is the order of multipacting.

Since the stalk has an impedance mismatch step, the calculations were carried out separately for these two parts. 

\begin{table}[h!]
	\caption{\label{tab:est}Analytical estimations in the stalk gap.}
	\begin{ruledtabular}
		\begin{tabular}{ccc}
			 & \multicolumn{2}{c}{Gun Voltage, kV} \\
			 \cline{2-3}
			Order of MP & Smaller Gap & Larger Gap \\ \hline
			1 & 143 & 342\\
			2 & 87 & 208 \\
			3 & 63 & 149\\
			4 & 49 & 116\\
			5 & 40 & 95\\
		\end{tabular}
	\end{ruledtabular}
\end{table}

One can see that analytical estimations of the threshold values for multipacting trajectories of the $1^{st}$ and $2^{nd}$ order in the stalk gap agree with the results of the numerical simulations, if we take into account the fact that two plane approach doesn't consider the real field distribution and phase of the electrons being released.

Another way to analytically estimate threshold levels of voltage is to use the diagram for multipacting bands in coaxial lines by E.~Somersalo \cite{Somersalo}. Our estimations using this chart showed that the first order two-point multipacting in the stalk can occur at 98 kV in the small and 411 kV in the big gap which is in a fair agreement with the estimations shown above.

\begin{figure}[h!]
	\includegraphics[width=1\linewidth]{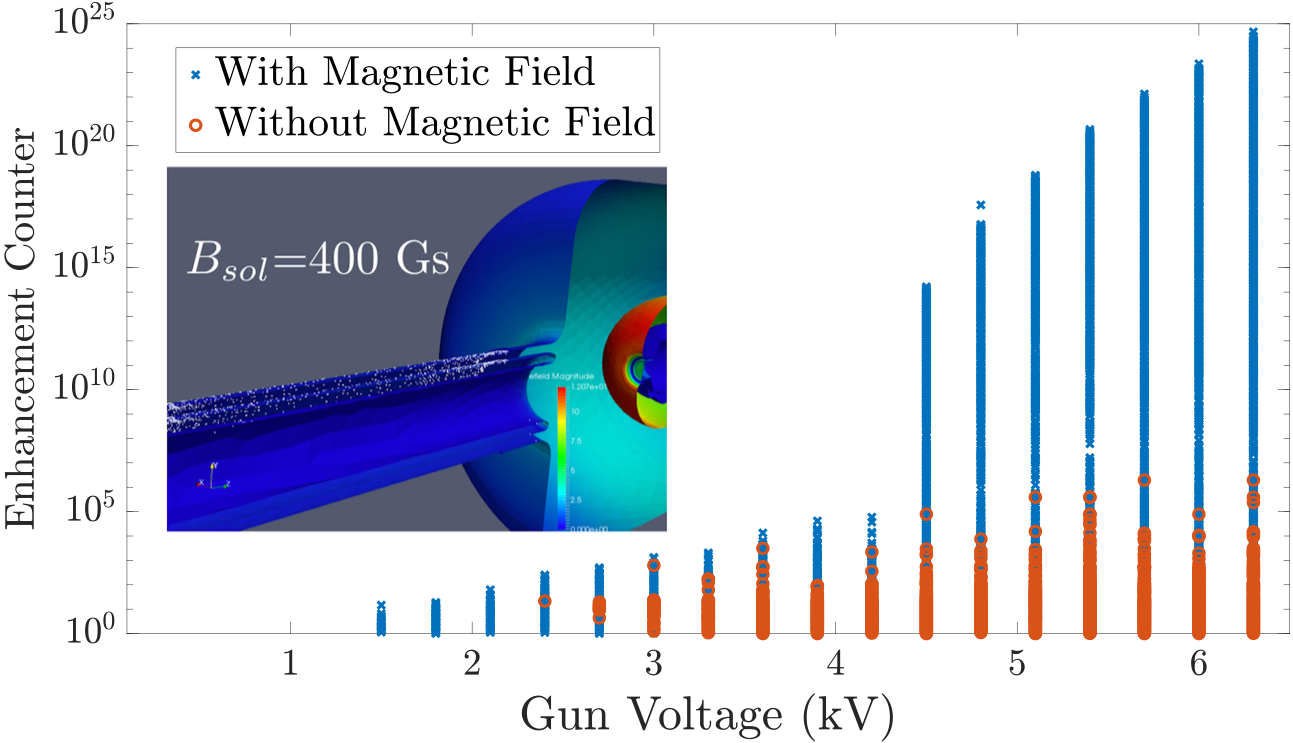}
	\caption{Enhancement Counter for the area of the FPC with (blue) and without (orange) external magnetic field.\label{fig:fpc}}
\end{figure}

A similar procedure of multipacting analysis was performed for the FPC area of the gun. This time primary electrons were emitted from a quarter of the FPC surface. First, the calculations were performed for low levels of accelerating voltage (0-7~kV) with and without external magnetic field. The magnetic field map was calculated using Poisson SUPERFISH \cite{poisson}, and corresponded to approximately 400~Gs on axis. It was confirmed that the introduction of an external magnetic field of the solenoid increases the strength of multipactor discharge in the gap between the FPC and the cavity wall (see Fig.~\ref{fig:fpc}).

\begin{figure}[h!]
	\includegraphics[width=1\linewidth]{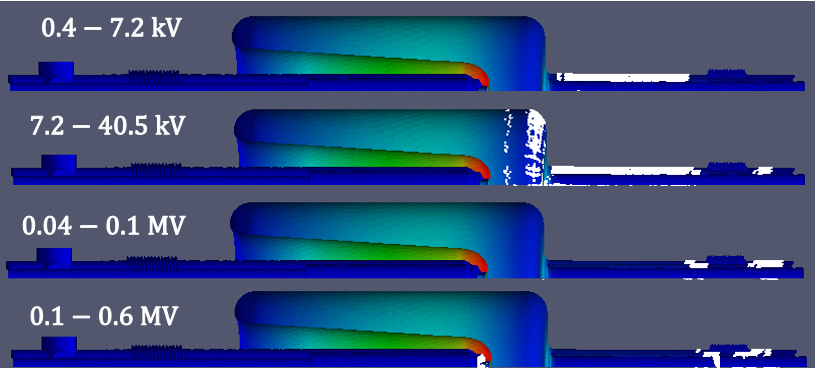}
	\caption{Areas of the gun affected by multipacting (shown in white). Primary electrons are emitted from the surface of the FPC.\label{traj2}}
\end{figure}

The resulting $EC$ function for the full range of operational voltage showed several peaks at low voltage in the gun, which corresponded to stable trajectories within the cavity body observed in the previous simulation, along with the $1^{st}$ order MP trajectories in the FPC. An external magnetic field was enabled for this simulation. One can see in Fig.~\ref{traj2}, that MP trajectories move along the FPC gap from cavity side toward the bellow when the gun voltage is increasing. At voltages above 500~kV, all of the stable trajectories are located in the stalk gap.

Based on the simulation results, we can conclude that the first multipacting trajectories within the cavity body occur at low accelerating voltages of about 7~kV, and are located at the front rounding and between the inner and outer conductors of the quarter-wave resonator. However, these trajectories are of the $8^{th}$ order, which can be easily processed during the conditioning. With voltage being increased, the trajectories in the front rounding become of a lower order, and around 30-40~kV only $1^{st}$ and $2^{nd}$ order trajectories are present in this area. This multipacting band ceases at 50~kV of accelerating voltage. The trajectories between the outer and inner conductors of the QWR move towards the back rounding of the cavity and become stable $1^{st}$ order trajectories around 200~kV. The FPC is found to be affected by multipactor discharge in a wider range of accelerating voltage, especially with the external field of the solenoid being applied. First, stable trajectories of the  $1^{st}$ and $2^{nd}$ order in the FPC gap appear at low voltages of about 5-7~kV and are located closer to the cavity. It is important to notice that some trajectories located between the nose of the cavity and its front rounding have been detected, but after a couple of RF cycles, they move towards the FPC gap and become $1^{st}$ and $2^{nd}$ order two-point multipacting trajectories between the FPC and the outer niobium pipe. Multipacting trajectories move away from the cavity along the FPC as voltage is increased, and are present in the bellow at a wide range of voltages starting from 100~kV. The discharge dies away in the area of the FPC for the voltages above 700~kV. Another area being affected by the discharge was found to be the stalk gap with stable trajectories being present at 150~kV and higher, up to the high end of the operational voltage.

\section{Experimental Results}

In the beginning of run 2017, multipacting behavior in the gun was studied separately during the RF system conditioning. Since vacuum activity is one of the signs of possible multipacting, we monitored pressure while varying the voltage in the cavity. The vacuum gauges are located in the laser cross (downstream of the cavity), the FPC and the cathode manipulator at the end of the stalk, which allowed us to judge which part of the system undergoes the multipacting. The measurements were performed without the cathode puck and for the solenoid field of about 400~Gs.

\begin{figure}[h!]
	\includegraphics[width=1\linewidth]{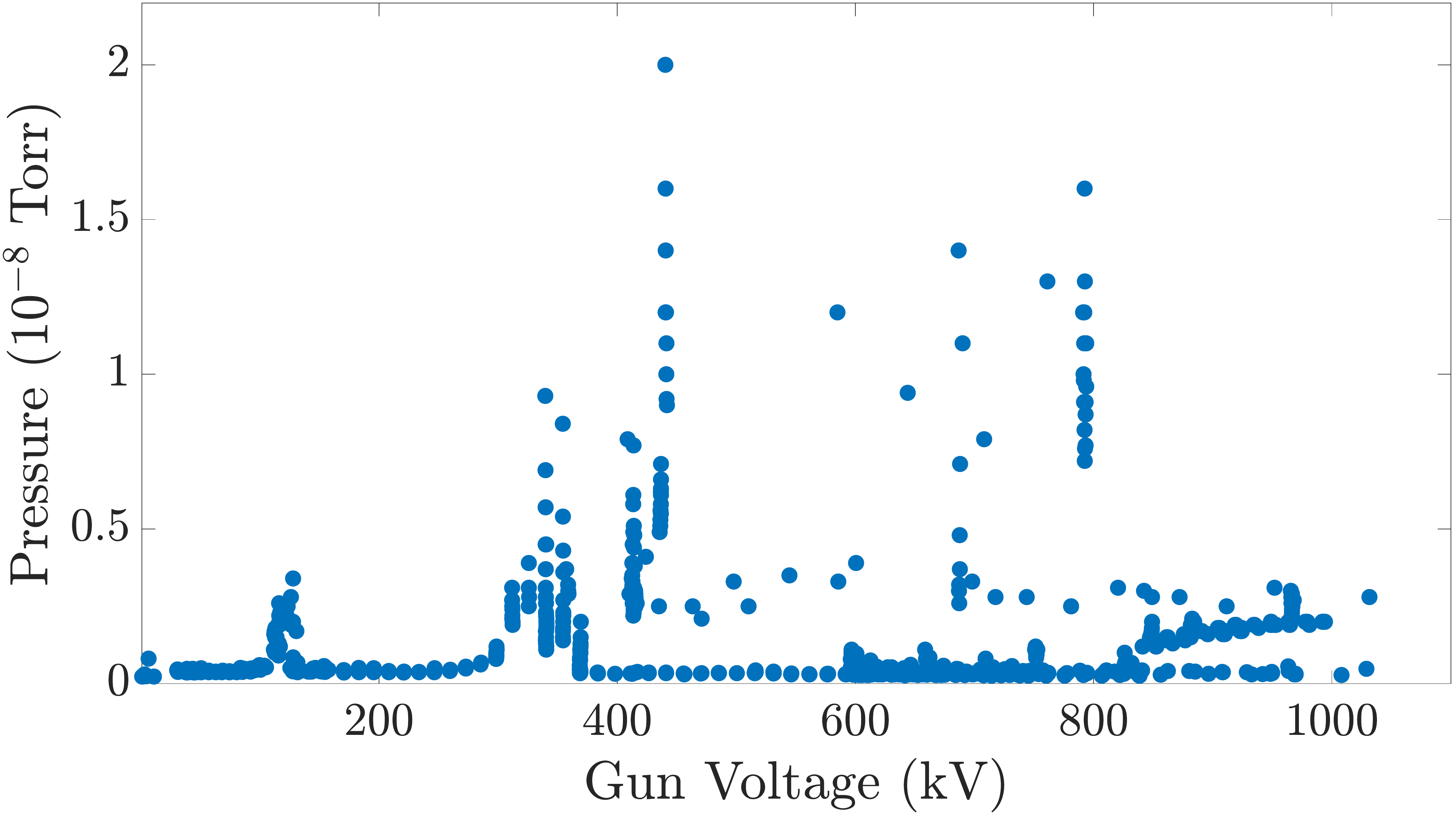}
	\caption{Pressure in the FPC during commissioning.\label{condit}}
\end{figure}

The CW conditioning results for the FPC are shown in Fig.~\ref{condit}. One can see that there was significant vacuum activity in the FPC at the gun voltage of about 120~kV, and at the voltages above 300~kV, which agrees with the predicted multipacting zones in the FPC gap. 
While it was possible to operate the gun stably above the 100-120~kV voltage, operation below 50~kV would always lead to the cavity being caught up on one of the low-voltage multipacting levels. Several multipacting levels were observed with vacuum activity detected in the FPC and the laser cross, with the latter being a sign of stable trajectories within the cavity body. These MP barriers occurred at about 4~kV, 30~kV and 40~kV---the levels which were observed during the previous run. In addition to increased vacuum pressure and cathode QE degradation, this was resulting in a typical ``zig-zag" pattern of the read-back voltage seen in Fig.~\ref{fig:Experiment}. Since we were not interested in the low voltage operation of our gun, we have never tried to operate at these levels and designed system to get through this zone as fast as possible.

During the CeC PoP commissioning, when the cathode puck was installed, the 40 kV MP level was the most challenging to overcome. To resolve this issue, a system startup script was written in order to capture the gun voltage above the dangerous multipacting zone as soon as it crossed the threshold. Before the power was delivered to the cavity, all the valves had to be closed, the laser turned off, and the solenoid current set to 0~A in order to avoid any possible enhancement of multipacting conditions. Then the FPC was brought to the position of maximal coupling with the cavity, and, with the phase-lock loop (PLL) turned on, the target frequency was checked at a low voltage ($\sim$ 1~kV). After setting the ``above the MP" threshold---typically 100~kV---an RF pulse with the full available power ($P_{1}$=4~kW), above the threshold ($P_{2}\sim$ 500~W), was sent in. During the full power pulse, low level RF (LLRF) system monitored the voltage for 400 milliseconds, and as soon as voltage rose above the threshold, the RF power was reduced to $P_{2}$. If the voltage did not rise above the threshold, the process was stopped with a request of operator assistance. After passing the MP barrier, the FPC was gradually moved to the operational position (defined by the operational frequency) and the power level was adjusted accordingly to maintain the desirable voltage in PLL mode. When the process was finished, the LLRF was switched to an IQ mode with flexible controls of gun voltage and phase.

This allowed us to reach operational gun voltage without tripping on the low level multipacting, while we still observed some vacuum activity in the FPC, cavity, and stalk during the start-up process. Even though the existing script significantly simplified reaching the desired operating voltage of the gun, the 40~kV multipacting level remained to be a problem in certain cases, and once the MP occurred, it was impossible to start-up the system right away. The only solution which helped in this situation, was to let the system ``rest" by leaving the gun off for about half an hour. After that, the script would bring the gun voltage to the operational regime without any problems. It is possible that this can be explained by the presence of the photocathode within the cavity body. Once the MP starts, the surface of the cathode would be affected by secondary electrons, which would lead to a deposit of active elements, such as Cs, on the walls of the cavity, increasing its SEY and making the MP worse. As mentioned above, multipacting conditions are strongly dependent on the quality of the surface, since all the processes are taking place within a thin layer on the surface, and any adsorbents can significantly increase the number of secondary electrons being released. The observed ``rest'' effect suggests that the multipacting induced Cs deposition may be transformed into a low-SEY material as a result of its reaction with residual gases. Further studies are needed to clarify this effect.

A few weeks into commissioning another multipacting barrier was found to be around 200~kV, which interrupted operation of the cavity, but didn't show any pressure changes in the vacuum gauges. This event can be explained by the stable trajectories located in the back rounding of the cavity (see Fig.~\ref{trajectories}d), so that the vacuum gauge in the laser cross located far away from the MP area couldn't detect the signal. 

Overall, performance of the gun was stable throughout the whole commissioning process, being interrupted by rare multipacting events, which could be easily overcome by turning the gun off for 30~minutes, and bringing it back on using the start-up script. It's important to notice that during the 3~months of operation, the cathode was replaced only once.

\section{Overcoming multipacting barrier}

\begin{figure}[h!]
	\includegraphics[width=1\linewidth]{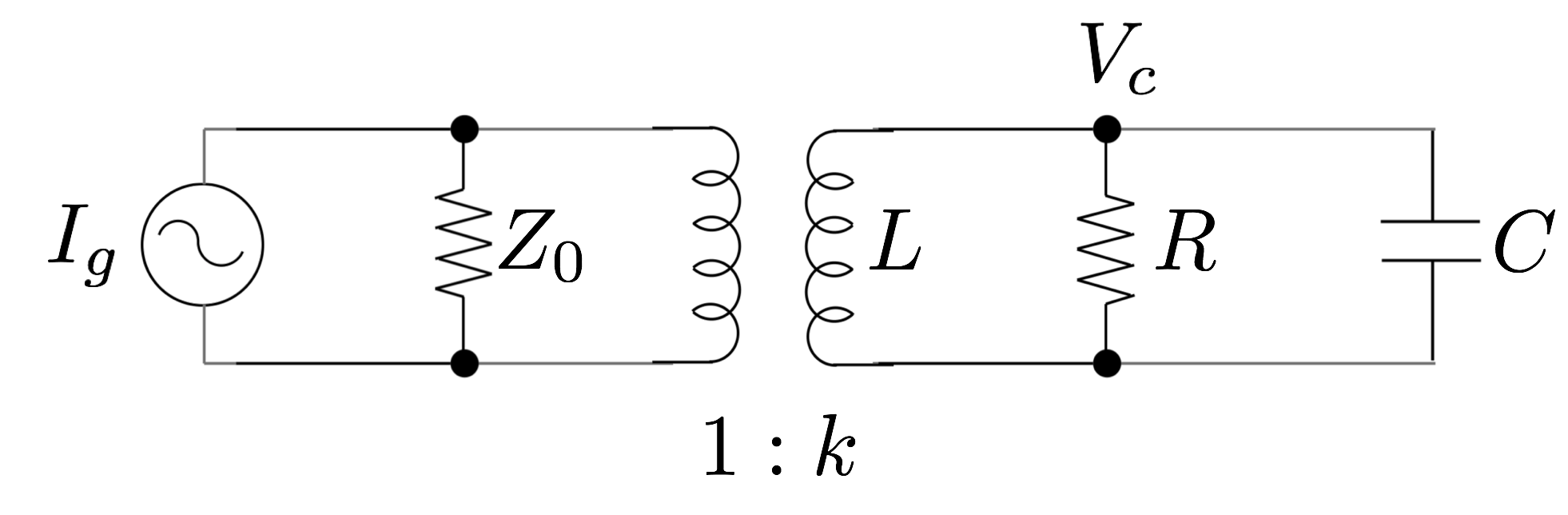}
	\caption{Equivalent circuit of a cavity with a power source.\label{fig:Equivalent1}}
\end{figure}

In order to overcome the multipacting zone, it is important to understand the influence of the coupling between the power source and the cavity. To do so, we used a well-known approach of modeling the interaction between a cavity and an RF transmitter as a circuit. The equivalent circuit is shown in Fig.~\ref{fig:Equivalent1}.

Using basic circuit theory and relating it to the known parameters of an RF cavity, voltage evolution in a cavity can be described by Eq.~\ref{eq:VoltageEvolutionText}, where $V_{0}$ is the maximum achievable voltage of the cavity, which is determined by the available input power and the coupling; $\tau$ and $\omega_{0}$ are the characteristic time and resonant frequency of a circuit correspondingly (see Appendix for the detailed discussion and derivations).

\begin{equation}
V_{c}=V_{0}\left(1-e^{-\dfrac{t}{2\tau}}\right)e^{i\omega_{0}t}.
\label{eq:VoltageEvolutionText}
\end{equation}

\begin{figure}[h!]
	\includegraphics[width=1\linewidth]{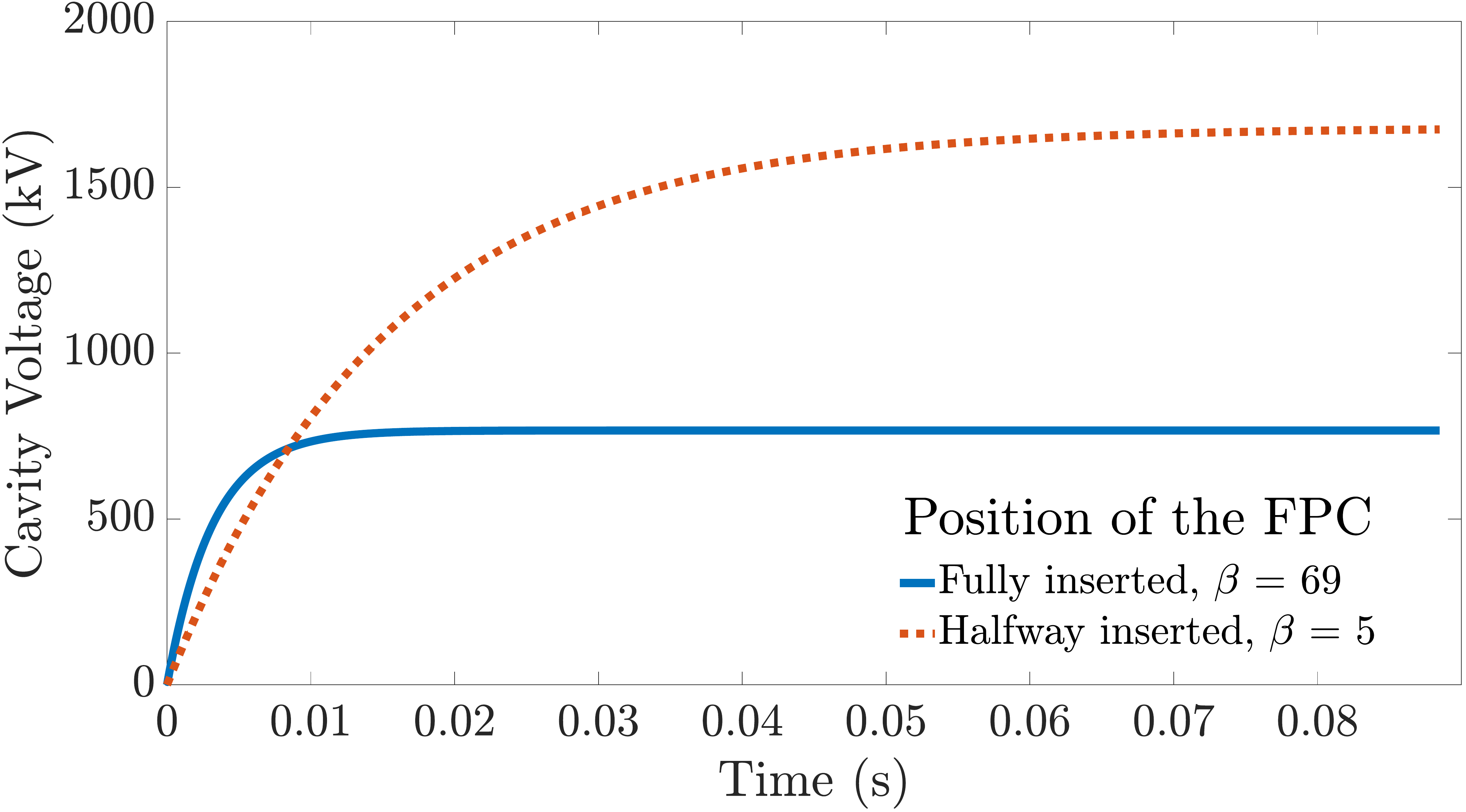}
	\caption{Cavity voltage as a function of time for various positions of the FPC.\label{Vc_vs_time}}
\end{figure}

The FPC of the cavity is designed to provide a 4.5~kHz tuning range with the total travel distance of 3~cm \cite{32}. Figure~\ref{Vc_vs_time} shows the filling of the cavity for different positions of the FPC. Filling time is crucial for surpassing multipacting zones at low voltages: if the filling time is shorter than the time it takes for the secondary electron avalanche to develop, multipacting should not interrupt the cavity operation.

To simulate the process of the cavity start-up, one needs to solve a system of two self-consistent differential equations: one describing the voltage evolution in a cavity with the presence of a multipactor discharge, and the second one is the equation of exponential growth of the secondary electron avalanche:

\begin{equation}
	\begin{cases}
	\dfrac{d|V_{c}|}{dt}=\frac{1}{2\tau}\left(|V_{0}|-|V_{c}|\right)-f_{0} \delta V_{mp}\dfrac{eN_{e}(t)}{2Q_{0}|V_{c}|}\omega_{0}R_{sh},\\
	\dfrac{dN_{e}}{dt}=\alpha(|V_{c}|)N_{e}.
	\end{cases}
\end{equation}

Energy losses due to the multipacting are determined by the number of the secondary electrons in the multipactor arc $N_{e}(t)$, voltage gained by the resonant particles $\delta V_{mp}$ and the RF parameters of the cavity---shunt impedance $R_{sh}$ and the quality factor $Q_{0}$.

\begin{figure}[h!]
	\includegraphics[width=1\linewidth]{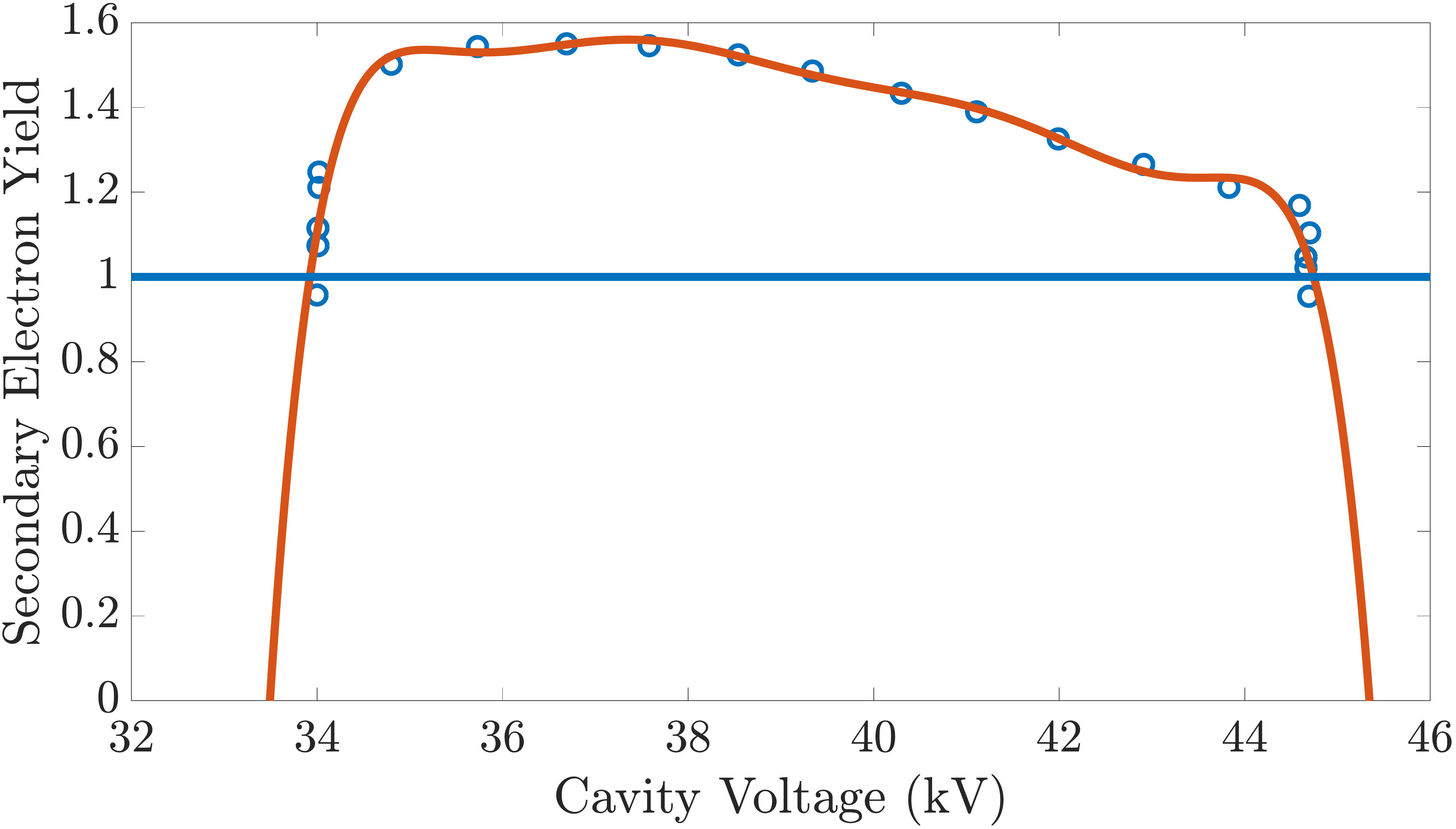}
	\caption{SEY coefficient of Nb based on the results of the $1^{st}$ order multipacting simulation using ACE3P.\label{SimSEY}}
\end{figure}

\begin{figure}[h!]
	\includegraphics[width=1\linewidth]{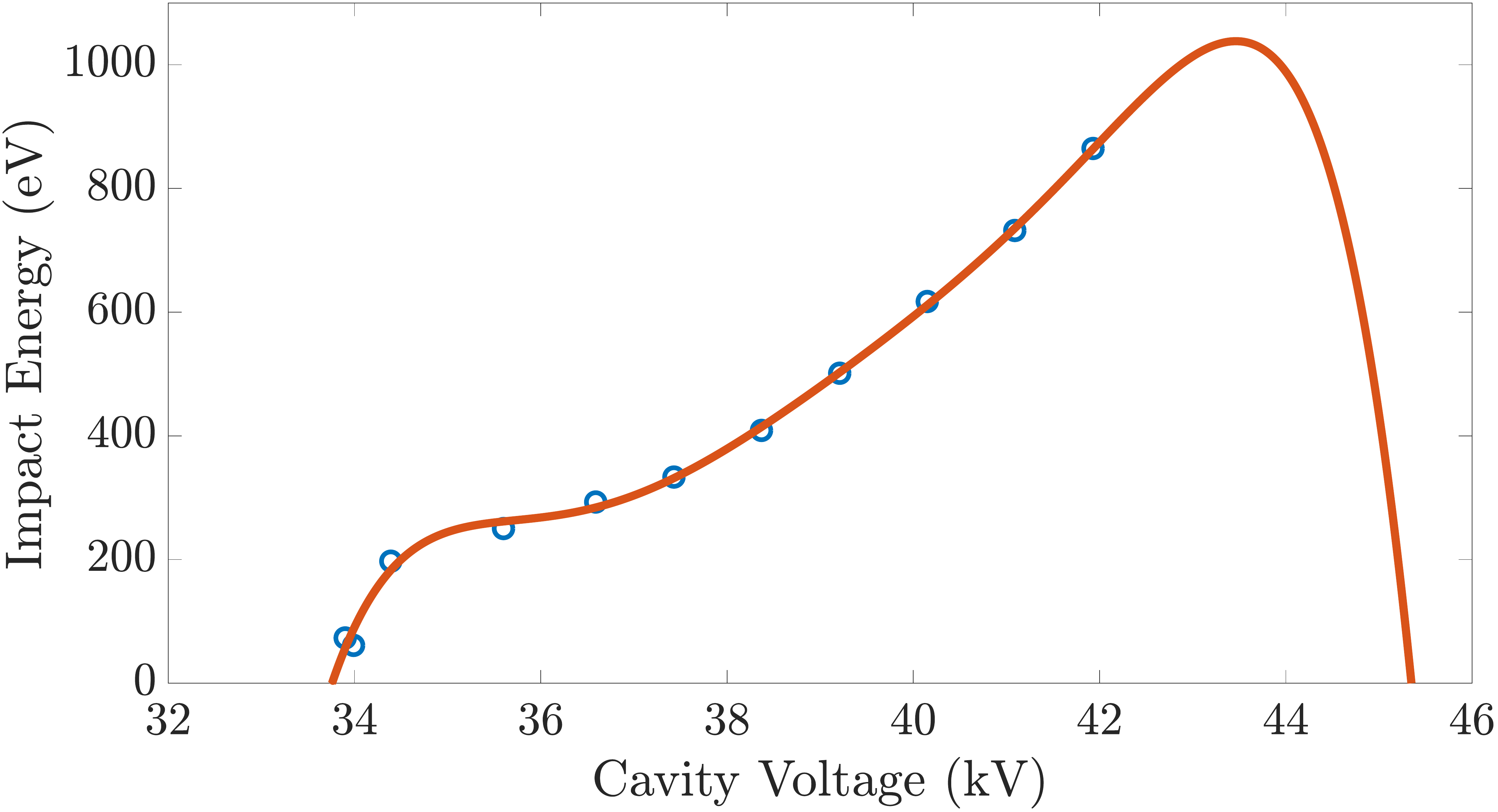}
	\caption{Impact energy of the secondary electrons based on the results of the $1^{st}$ order multipacting simulation using ACE3P.\label{SimEnergy}}
\end{figure}

In order to take into account geometrical conditions for the multipacting discharge within the cavity, we used the simulation results for the $1^{st}$ and $2^{nd}$ order multipacting at the low voltage levels of about 20-40~kV. The data for the Enhancement Counter and the impact energy of the secondary electrons at the multipacting zones were fitted and used in our simulations to determine the SEY coefficient and the power consumed by the multipacting. Fig.~\ref{SimSEY} and \ref{SimEnergy} show the fits of the SEY coefficient and impact energy dependencies used for the simulations of the $1^{st}$ order multipacting in the front rounding of the cavity. 

The simulations were performed for two different SEY coefficients (for niobium and copper), various positions of the FPC relative to the cavity, and different maximum available forward power. We observed the evolution of the cavity voltage and the number of secondary electrons within the cavity volume.

\begin{figure}[h!]
	\includegraphics[width=1\linewidth]{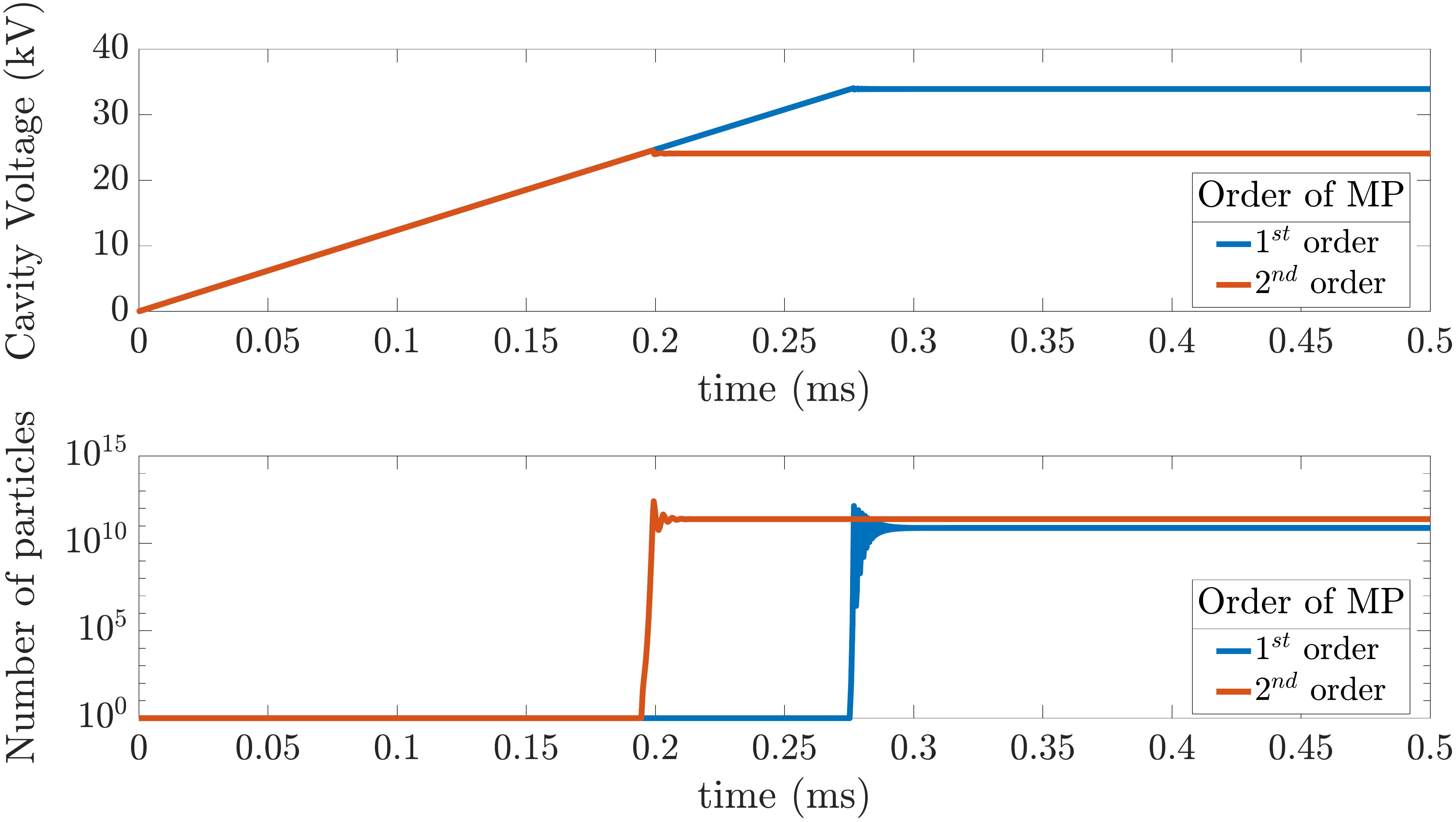}
	\caption{Cavity voltage and the number of secondary electrons as a function of time: $1^{st}$ and $2^{nd}$ order multipacting with the FPC being incerted half-way in, maximal aviliable power of 4~kW, and SEY for Nb.\label{MP_FPC_mid}}
\end{figure}

Figure~\ref{MP_FPC_mid} shows the results of simulation for the FPC being inserted half-way in with the maximal available power of 4~kW and SEY coefficient for niobium. As expected, when turning on the cavity, voltage grows linearly with time until it reaches the lower barrier of the multipacting zone. The number of secondary electrons starts to grow exponentially and the avalanche consumes the input power, which leads to the oscillations of the cavity voltage around the lower multipacting bound.

\begin{figure*}
\begin{minipage}[t]{0.48\textwidth}
\includegraphics[width=\textwidth]{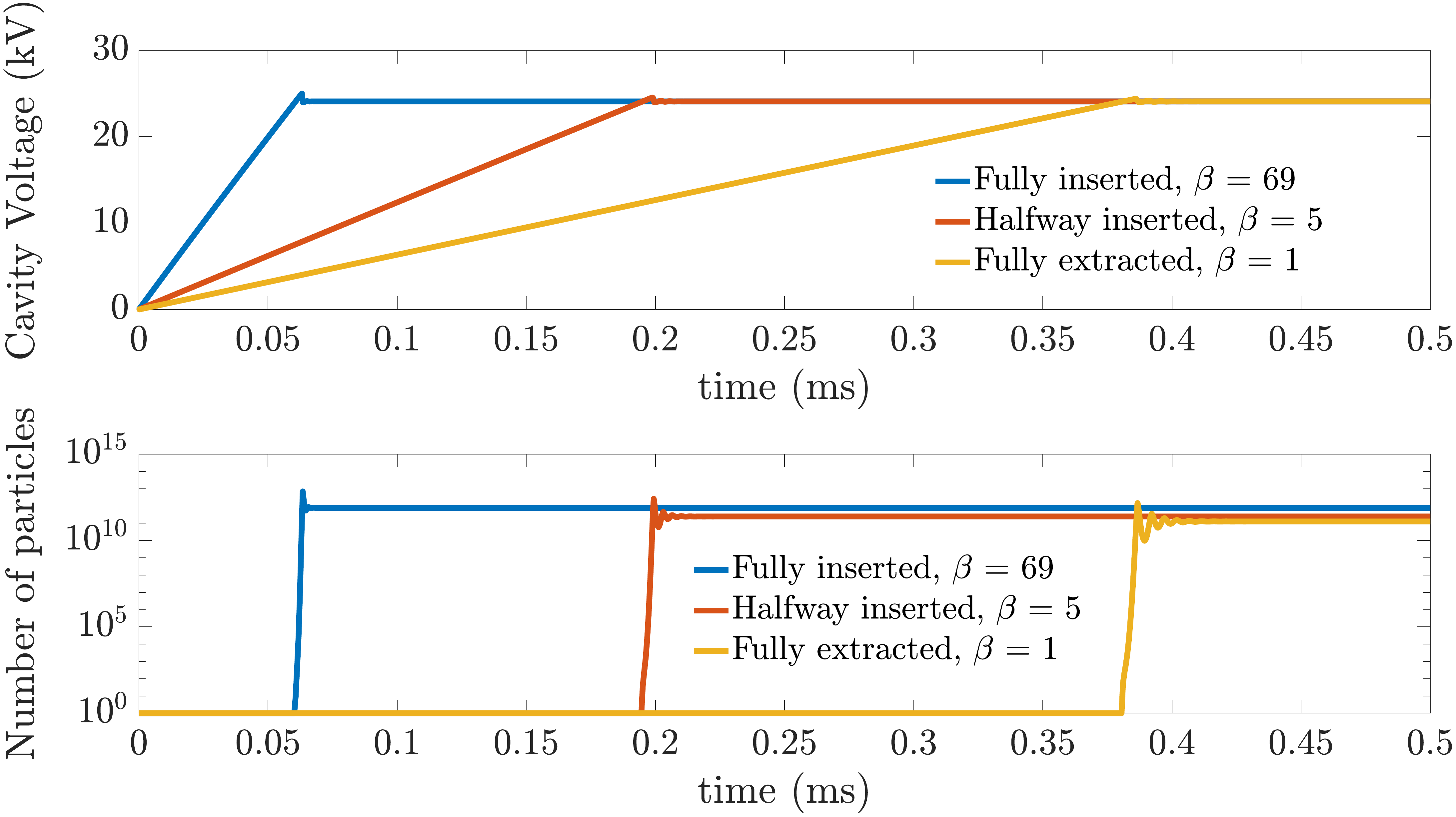}\\
(a)
\end{minipage}
\begin{minipage}[t]{0.48\textwidth}
\includegraphics[width=\textwidth]{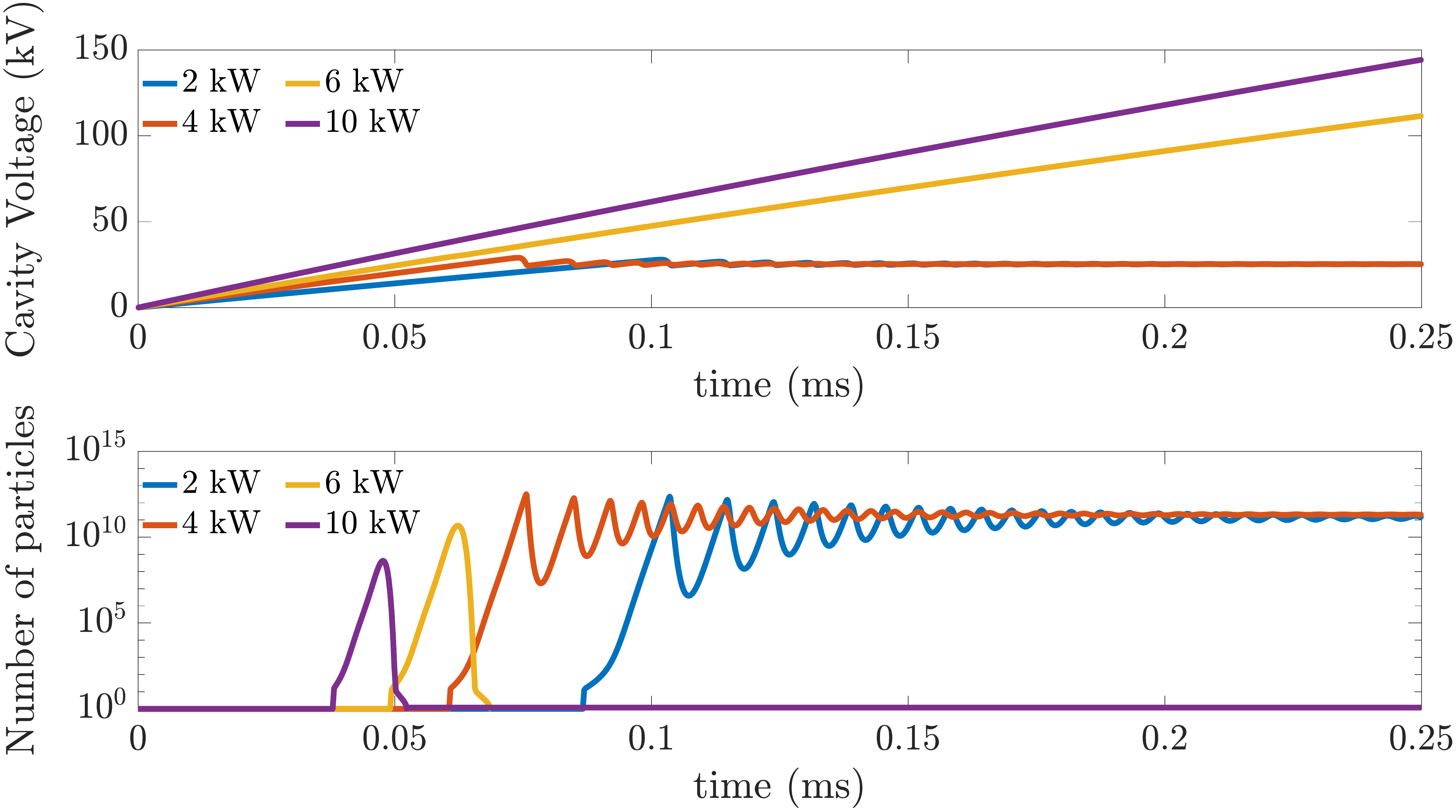}\\
(b)
\end{minipage}
\begin{minipage}[t]{0.48\textwidth}
\includegraphics[width=\textwidth]{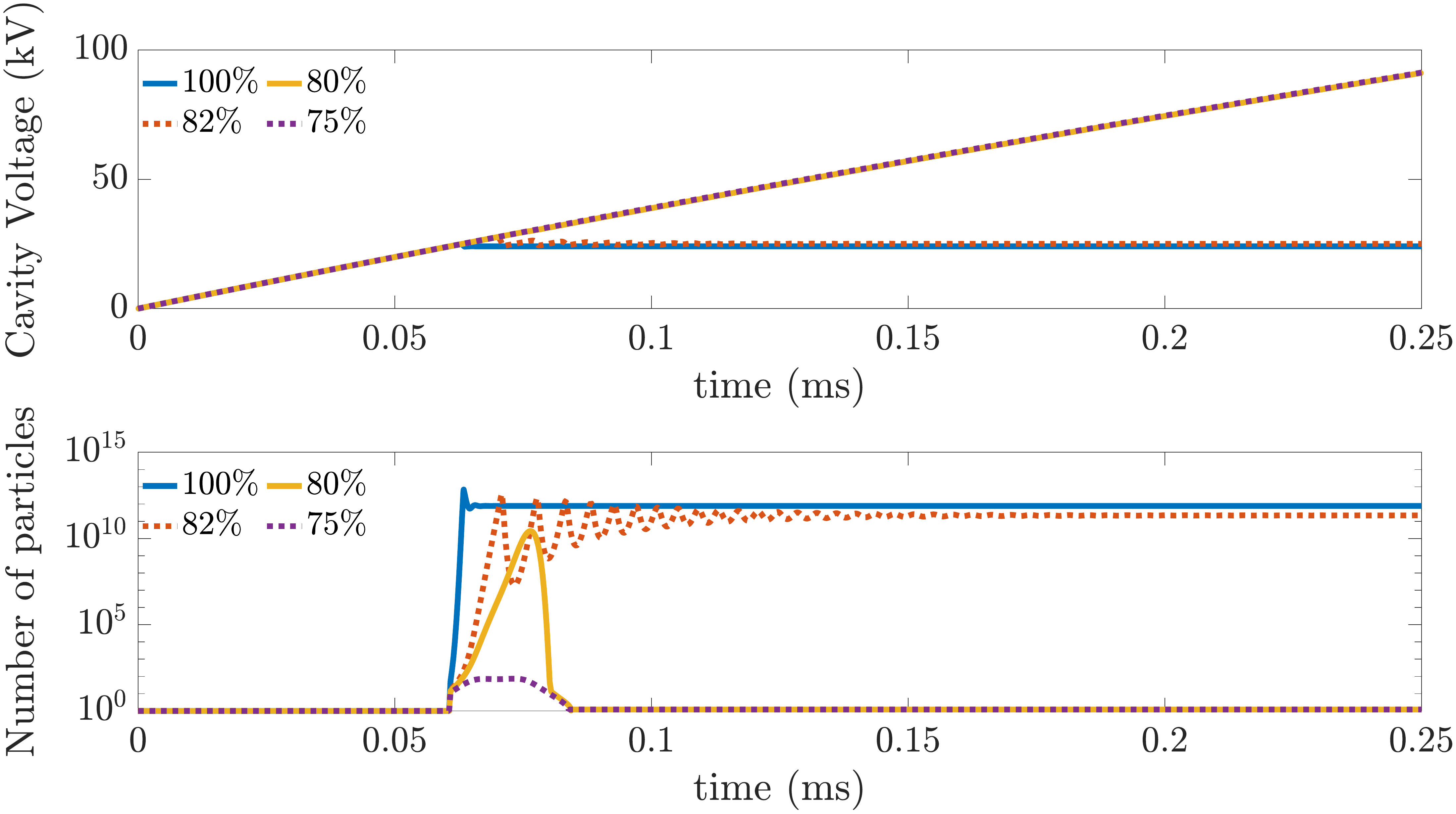}\\
(c)
\end{minipage}
\begin{minipage}[t]{0.48\textwidth}
\includegraphics[width=\textwidth]{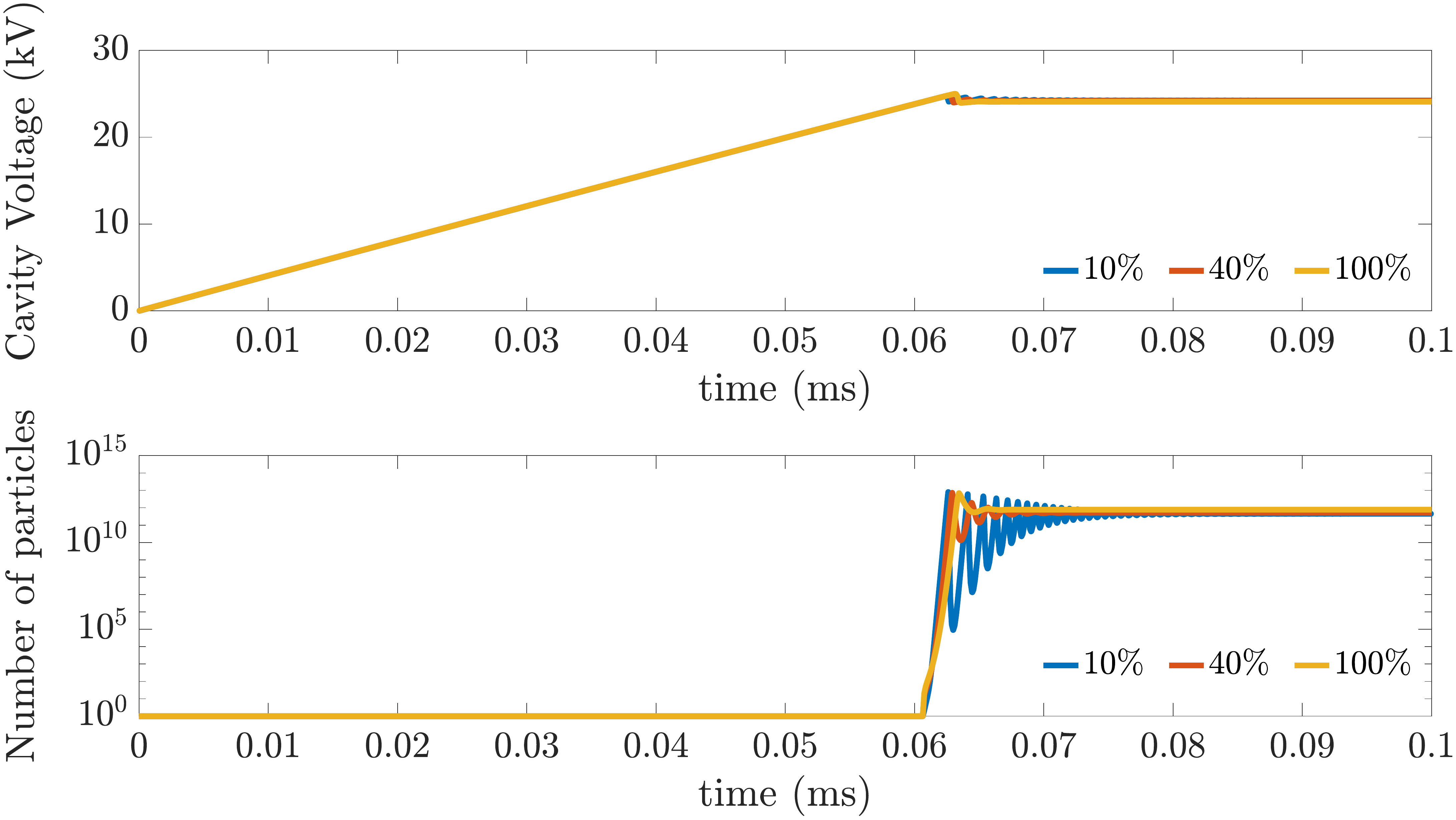}\\
(d)
\end{minipage}
\caption{Cavity voltage and the number of secondary electrons as a function of time with the presence of the $2^{nd}$ order MP: (a) SEY for Nb, 4~kW input power and various positions of the FPC; (b) fully inserted FPC, SEY for Nb and various values of the input power; (c) fully inserted FPC, 4~kW input power and different uniform scaling of the SEY for Nb; (d) fully inserted FPC, 4~kW input power,  SEY for Nb for various widths of the MP zone.}
\label{MPcode_parameters}
\end{figure*}

Setting the coupling coefficient to the one corresponding to the maximal insertion of the FPC did not allow us to pass the multipacting barrier, which can be explained by several approximations used in the simulation. First, the actual SEY coefficient of the cavity surface is unknown and might differ significantly from the default curve used in the simulations. Even though we used the results from ACE3P in order to account for the geometrical resonant conditions, the width of the multipacting zone might vary in reality, which can possibly make it easier to pass the multipacting barrier. And, finally, we supposed that none of the secondary electrons get lost during the process, and all of the emitted particles are released in the phase satisfying the resonant conditions. Introducing several scaling parameters in order to accurately approximate the conditions described above, we can demonstrate the influence of these parameters on the multipactor discharge in the cavity (see an example for the $2^{nd}$ order MP in Fig.~\ref{MPcode_parameters}). 

\begin{figure}[h!]
	\includegraphics[width=1\linewidth]{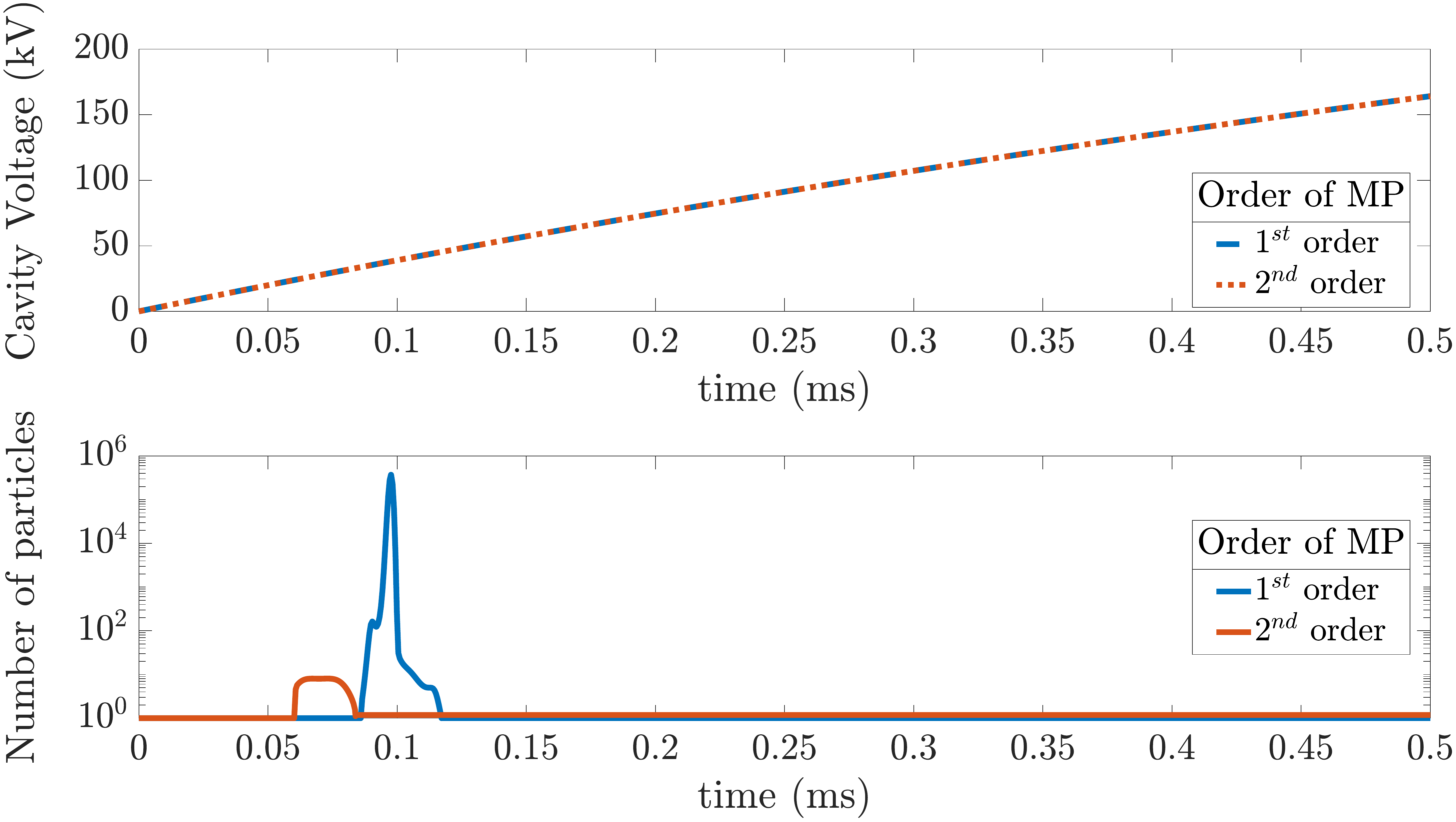}
	\caption{Cavity voltage and the number of secondary electrons as a function of time: $1^{st}$ and $2^{nd}$ order multipacting with the maximum coupling, aviliable power of 4~kW, and adjusted SEY for Nb.\label{fig:MP_passed}}
\end{figure}


Figure.~\ref{fig:MP_passed} shows a successful passage of the multipacting events (both $1^{st}$ and $2^{nd}$ order) with the maximum coupling and 4~kW available power of the RF transmitter. The result was achieved by reducing the SEY uniformly up to 72\% of the original value with 90\% of the particles being involved in the MP process. One can see that the growth of the secondary electrons is still present within the cavity body, however, since the position of the FPC is providing the maximum coupling, we cross over the resonant conditions before the electron avalanche can form and bring the cavity down to the multipacting level.

Such an exercise clearly shows that strong coupling and clean surface of the cavity walls are the key components to the successful passing of the low level multipacting.

\section{Conclusion}

While SRF guns had been considered as a potential breakthrough in the area of high brightness electron sources, there was a very important question about compatibility of SRF cavities and high QE (and high SEY) photocathodes. The problem is that the deposition of active elements from high QE photocathodes, such as Cs, on the surface of a cavity, can lead to higher SEY, making the cavity more vulnerable to multipacting and affect operation of an SRF cavity. On the other side, dark current, back-bombardment and multipacting in SRF cavities could affect the QE of the cathodes. These issues must be considered in the early stages of SRF gun design and development. However, our theoretical and experimental studies, along with our simulations, showed that there is a solution for this challenge. 

Strong coupling plays a crucial role in overcoming any low level MP barriers: time for crossing the multipacting zone must be significantly shorter than the time it takes for the secondary electron avalanche to develop in order to pass multipacting. Providing the maximal coupling and full available power of an RF transmitter while bringing an SRF gun to the operational level, and supporting cleanliness of the cavity walls are the key components of avoiding multipactor discharge.

\begin{acknowledgments}
The authors would like to thank Lixin Ge from SLAC and Tianmu Xin from BNL for helpful discussions and useful tips on using ACE3P.

This research used resources of the National Energy Research Scientific Computing Center, which is supported by the Office of Science of the U.S. Department of Energy under Contract No.DE-AC02-05CH11231.

Work is supported by Brookhaven Science Associates, LLC under Contract No. DEAC0298CH10886 with the U.S. Department of Energy, DOE NP office grant DE-FOA- 0000632, and NSF grant PHY-1415252.
\end{acknowledgments}

\appendix
\label{appendix}
\section{Connecting the Equivalent Circuit Parameters with the Characteristic Quantities of the Cavity}
\begin{figure}[h!]
	\includegraphics[width=1\linewidth]{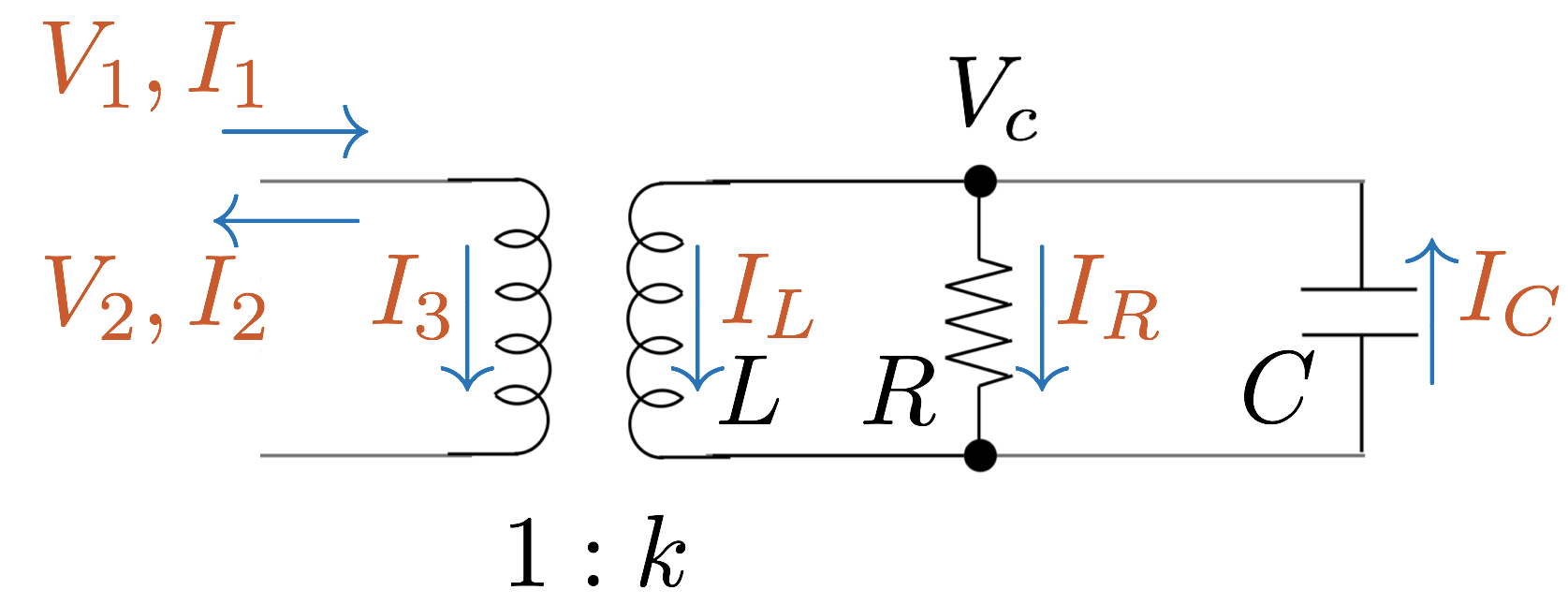}
	\caption{Definition of the currents and voltages in the equivalent circuit of the cavity with a power supply. \label{fig:Equivalent2}}
\end{figure}

In order to describe the interaction between the cavity and the power source, we will represent the system as an equivalent circuit shown earlier in the paper (see Fig.~\ref{fig:Equivalent1}), and we will define all the currents and voltages going through the system including forward $I_{1}$, $V_{1}$ and reflected $I_{2}$, $V_{2}$ waves in the way shown in Fig.~\ref{fig:Equivalent2}.

Starting with the basic principles, we will derive the relations between the parameters of the equivalent circuit and an RF cavity. The relationships can be found in any fundamental book on RF technology (see \cite{31}), but we will provide main equations and definitions for the completeness of the derivation. The stored energy in the RLC circuit shown above can be expressed as:

\begin{equation}
U=\frac{C|V_{c}|^2}{2}.
\end{equation}

The dissipated power in the cavity walls $P_{c}$ is equivalent to the resistive losses in the circuit:

\begin{equation}
P_{c}=\frac{|V_{c}|^2}{2R}.
\end{equation}

Intrinsic quality factor of an RF cavity $Q_{0}$ can be written in terms of the circuit parameters as follows:

\begin{equation}
Q_{0}=\frac{\omega_{0}U}{P_{c}}=\sqrt{\frac{C}{L}}R.
\end{equation}

The power source in the equivalent circuit representation can be modeled as an ideal current source with a shunt impedance $Z_{0}$ as shown in Fig.~\ref{fig:Equivalent1}. The cavity being connected to the RF transmitter sees the additional admittance $\frac{1}{Z_{0}k^2}$ of the transmission line through the transformer with a coupling coefficient $k$, which gives us an expression for the losses through the coupler $P_{e}$ (the power leaking back out the input coupler):

\begin{equation}
P_{e}=\frac{1}{Z_{0}k^2}\frac{|V_{c}|^2}{2}.
\end{equation}

Knowing that, the external quality factor $Q_{e}$ can be also expressed in terms of the equivalent circuit parameters:

\begin{equation}
Q_{e}=\frac{\omega_{0}U}{P_{e}}=\sqrt{\frac{C}{L}}Z_{0}k^2.
\end{equation}

The coupling coefficient $\beta$ is defined as a ratio between the power dissipated in the external circuits and the power lost within the cavity walls, or, alternatively, it can be written as:

\begin{equation}
\beta=\frac{Q_{0}}{Q_{e}}=\frac{1}{k^2}\frac{R}{Z_{0}}.
\end{equation}

Finally, one has to remember that the RF definition of the shunt impedance is twice the resistance of the equivalent circuit $R_{sh}=2R$, which gives us coupling coefficient of the transformer in terms of the RF cavity parameters:

\begin{equation}
\label{eq:n}
k=\sqrt{\frac{1}{\beta}\frac{R}{Z_{0}}}=\sqrt{\frac{R_{sh}}{2\beta Z_{0}}}.
\end{equation}

\section{Equation for the Voltage Evolution}

With the voltages and currents in the system defined earlier (see Fig.~\ref{fig:Equivalent2}), we will derive the equation for the evolution of the cavity voltage $V_{c}$. One can start by finding all the currents and voltages going through the equivalent circuit: 

\begin{equation}
    \begin{cases}
	I_{R}=-\dfrac{V_{c}}{R},\hspace{0.5cm} I_{L}=-C\dfrac{dV_{c}}{dt}-\dfrac{V_{c}}{R},\\
    I_{3}=\dfrac{V_{1}-V_{2}}{Z_{0}}=I_{1}-I_{2},\hspace{0.5cm} I_{R}+I_{L}=-\dfrac{dQ}{dt},\\
    V_{1}=Z_{0}I_{1},\hspace{0.5cm} V_{2}=Z_{0}I_{2},\hspace{0.5cm} V_{c}=\dfrac{Q}{C}.
	\end{cases}
\end{equation}

\begin{equation}
V_{c}=L\left(\dfrac{dI_{L}}{dt}+\dfrac{1}{k}\dfrac{dI_{3}}{dt}\right)=k(V_{1}+V_{2}).
\end{equation}

Using $V_{c}$ as the main variable and eliminating all of the currents and their derivatives gives:

\begin{equation}
\dfrac{d^{2}V_{c}}{dt^2}+\dfrac{1}{\tau}\dfrac{dV_{c}}{dt}+\omega_{0}^2V_{c}=\dfrac{1}{T}\dfrac{dV_{1}}{dt}.
\label{eq:CircuitVoltage}
\end{equation}

In the Eq.~\ref{eq:CircuitVoltage}, we introduced following parameters in order to simplify the resulting equations:

\begin{equation}
	\begin{cases}
	\dfrac{1}{\tau}=\dfrac{1}{CR}+\dfrac{1}{k^2CZ_{0}}=\dfrac{\omega_{0}}{Q_{L}},\\
	\omega_{0}^2=\dfrac{1}{LC},\\
	 \dfrac{1}{T}=\dfrac{2}{kZ_{0}C}.
	\end{cases}
\end{equation}

First we can take look at the stationary solution of the Eq.~\ref{eq:CircuitVoltage}. Let us denote the amplitude of a cavity voltage to be $V_{0}$, and the amplitude of the forward wave $V_{1}$ to be $V_{g}=\sqrt{2P_{g}Z_{0}} $ with $P_{g}$ standing for the available generator power:

\begin{equation}
	\begin{cases}
	V_{1}=V_{g}e^{i\omega t}, \\
    V_{c}=V_{0}e^{i\omega t}. \\
	\end{cases}
\end{equation}

After plugging this solution into the differential equation, one can find that at the resonance when $\omega = \omega_{0}$, we can obtain an expression for the maximal achievable voltage in the cavity $V_{0}$ which is determined by the available forward power $P_{g}$ and the coupling coefficient $\beta$:

\begin{equation}
	V_{0}=\dfrac{\tau}{T}V_{g}=\dfrac{2}{1+\beta}\sqrt{\beta R_{sh}P_{g}}.
\end{equation}

We will omit the process of solving the Eq.~\ref{eq:CircuitVoltage}, and provide the resulting solution which describes the evolution of the cavity voltage with the maximal achievable voltage $V_{0}$ :

\begin{equation}
V_{c}=V_{0}\left(1-e^{-\dfrac{t}{2\tau}}\right)e^{i\omega_{0}t}.
\label{eq:VoltageEvolution}
\end{equation}

For the initial build up of the cavity voltage ($t\ll\tau$), which will be of the main interest for the low level multipacting simulations, the Eq.~\ref{eq:VoltageEvolution} can be expanded:

\begin{equation}
V_{c}\cong V_{0}\dfrac{t}{2\tau}e^{i\omega_{0}t}=\dfrac{t}{2T}V_{g}e^{i\omega_{0}t}=\sqrt{\beta \dfrac{R}{Z_{0}}}\dfrac{V_{g}t}{RC}e^{i\omega_{0}t}.
\label{eq:VoltageEvolutionExpanded}
\end{equation}

\begin{equation}
P_{c}=\dfrac{|V_{c}|^2}{R_{sh}}=\beta \dfrac{R}{R_{sh}} \dfrac{V_{g}^2}{Z_{0}}\left(\dfrac{t}{RC}\right)^2=\beta P_{g} \left(\dfrac{t}{RC}\right)^2.
\label{eq:PowerFromExpantion}
\end{equation}

Based on the Eq.~\ref{eq:PowerFromExpantion} one can make a conclusion, that strong coupling is the key for the fast growth of the cavity voltage: to be specific, the square root of the product of the coupling $\beta$ and the generator power $P_{g}$ is what drives the growth of the voltage.

\section{Adding Multipacting}

Evolution of the stored energy within the cavity $W$ without the multipacting can be derived from the known equation of the cavity filling (see Eq.~\ref{eq:VoltageEvolution}):

\begin{equation}
W(t)=W_{0}\left(1-e^{-\dfrac{t}{2\tau}}\right)^{2},
\label{eq:EnergyEvolution}
\end{equation}

with $W_{0}=\dfrac{C|V_{0}|^2}{2}$.

This can be used to derive following differential equation, which describes the dependence of the stored energy on time:

\begin{equation}
\dfrac{dW(t)}{dt}+\dfrac{W(t)}{\tau}=\dfrac{1}{\tau} \sqrt{W(t)}\sqrt{W_{0}}.
\label{eq:EnergyDiffEq}
\end{equation}

Now we can take into account energy losses due to the multipacting discharge, which gives following:

\begin{equation}
\dfrac{dW}{dt}+\dfrac{W}{\tau}=\dfrac{1}{\tau} \sqrt{W}\sqrt{W_{0}}-\dfrac{2e\Delta V_{mp}f_{0}N_{e}(t)}{2n-1}.
\label{eq:EnergyDiffEqWithMP}
\end{equation}

In the Eq.~\ref{eq:EnergyDiffEqWithMP} $N_{e}(t)$ is number of electrons in the multipactor arc, $e\Delta V_{mp}$ is energy gained by a multipacting electron in a half of the oscillation period, $f_{0}=$ 113~MHz is resonant frequency of the cavity, and $n$ is order of multipacting---a number of half-periods needed for electron to reach the surface after emission.

In terms of the cavity voltage, Eq.~\ref{eq:EnergyDiffEqWithMP} will become:

\begin{equation}
\dfrac{d|V_{c}|}{dt}=\dfrac{1}{2\tau}\left(|V_{0}|-|V_{c}|\right)-\dfrac{2ef_{0}N_{e}(t)\Delta V_{mp}}{(2n-1)C|V_{c}|}.
\label{eq:VoltageDiffEqWithMP}
\end{equation}

Let us introduce new parameters to simplify further derivations: 

\begin{equation}
	\begin{cases}
		 \delta V_{mp}=\dfrac{2 \Delta V_{mp}}{2n-1},\\
		 r=\dfrac{k^{2}Z_{0}R}{k^{2}Z_{0}+R},\\ 
		 \dfrac{C}{\tau}=\dfrac{1}{R}+\dfrac{1}{k^{2}Z_{0}}=\dfrac{1}{r}.
	\end{cases}
\end{equation}

Condition for the multipacting to overpower the power transmitted will be when the change in the voltage becomes negative:

\begin{equation}
\dfrac{d|V_{c}|}{dt}=\dfrac{1}{2\tau}\left(|V_{0}|-|V_{c}|\right)-\dfrac{2ef_{0}N_{e}(t)\Delta V_{mp}}{(2n-1)C|V_{c}|}\leq 0,
\end{equation}

\begin{equation}
f_{0}e\delta V_{mp} N_{e}(t)\geq \dfrac{1}{2}\dfrac{|V_{c}|(|V_{0}|-|V_{c}|)}{r}.
\end{equation}

Now we can add the evolution equation for the secondary electrons within the cavity to account for the growth of the electron avalanche with time:

\begin{equation}
\dfrac{dN_{e}}{dt}=\alpha N_{e}.
\label{eq:SecElEvolution}
\end{equation}

In the Eq.~\ref{eq:SecElEvolution}, $\alpha$ is the rate of the exponential growth of the secondary electrons, and it is related to the SEY coefficient $\delta$ and order of multipacitng $n$. In order to obtain this relationship, one can use Eq.~\ref{eq:Nexp} to find the number of secondary particles produced by a single initial electron after one impact, which, by definition, is equal to the SEY coefficient $\delta$. The time it takes for an MP electron to reach the wall depends on the order of multipacting $n$ and always equals to an odd number of half-periods of the RF field $\Delta t=\dfrac{(2n-1)}{2f_{0}}$. Solving for $\alpha$ gives us following relationship between $\alpha$ and $\delta$:

\begin{equation}
\alpha(|V_{c}|)=\dfrac{2 f_{0} ln(\delta(|V_{c}|))}{(2n-1)}.
\label{eq:alpha}
\end{equation}

Altogether we obtain a self-consistent set of differential equations for modeling the voltage evolution within the cavity in the presence of multipacting:

\begin{equation}
	\begin{cases}
	\dfrac{d|V_{c}|}{dt}=\frac{1}{2\tau}\left(|V_{0}|-|V_{c}|\right)-f_{0} \delta V_{mp}\dfrac{eN_{e}(t)}{2Q_{0}|V_{c}|}\omega_{0}R_{sh},\\
	\dfrac{dN_{e}}{dt}=\alpha(|V_{c}|)N_{e}.
	\end{cases}
\end{equation}

\bibliography{Draft_aps}

\end{document}